\documentclass[twocolumn]{webofc}
\def\ba{\begin{eqnarray}}
\def\ea{\end{eqnarray}}

\usepackage{xcolor}

\usepackage[varg]{txfonts}   % Web of Conferences font
\begin{document}
\title{Quark model calculations of transition form factors 
at high photon virtualities}

\hspace{14cm} \mbox{\bf LFTC-19-12/50}

\author{\firstname{G.~Ramalho} 
\inst{1}%\fnsep\thanks{\email{gilberto.ramalho@universidadecruzeirodosul.edu.br}}
        % etc.
}

\institute{Laborat\'orio de 
F\'{i}sica Te\'orica e Computacional -- LFTC,
Universidade Cruzeiro do Sul/Universidade Cidade de  S\~ao Paulo,\\
 01506-000,   S\~ao Paulo, SP, Brazil}

\abstract{We present calculations 
of  $\gamma^\ast N \to N^\ast$
transition form factors, where $N$ is the nucleon 
and $N^\ast$ is a nucleon resonance,
based on a covariant quark model.
Our main focus is at high photon virtualities (large $Q^2$) 
where the valence quark degrees of freedom 
dominate the contributions to the transition form factors
and helicity amplitudes.
In that regime, the quark model estimates 
can be compared with the available data,
particularly with the Jefferson Lab data 
at intermediate and large momentum transfer ($Q^2 >2$ GeV$^2$).
The main focus is on the $\Delta(1232)3/2^+$, 
$N(1440)1/2^+$,  $N(1535)1/2^-$ and  $N(1520)3/2^-$ resonances,
but estimates for other higher mass resonances are also discussed.}
\maketitle

\section{Introduction}
\label{secIntro}

In the last two decades important information about 
the electromagnetic structure of the nucleon 
and the nucleon resonances  have been collected 
in modern facilities such as Jefferson Lab (JLab), MAMI, MIT-Bates and  
ELSA, among others~\cite{NSTAR,Aznauryan12,Burkert04,Drechsel07,NSTAR2017}.
Helicity amplitudes and form factors associated with the 
$\gamma^\ast N \to N^\ast$ transitions have been measured 
for resonances $N^\ast$ with masses up to 2 GeV for values of $Q^2$ in 
the range $Q^2= 0$-- 8 GeV$^2$~\cite{NSTAR,MokeevDatabase},
where $Q^2= -q^2$, and $q$ is the momentum transfer.
With the JLab-12 GeV upgrade, we expect to probe 
even larger values of $Q^2$~\cite{NSTAR,NSTAR2017}.

To interpret the new data above $Q^2=2$ GeV$^2$, theoretical models 
based on relativity are fundamental.
Models which include valence quark degrees of freedom
are preferable, since those are the 
expected dominant degrees of freedom at large $Q^2$~\cite{NSTAR,Aznauryan12,Burkert04,Drechsel07}.
These models can be used to make predictions for 
transition form factors at large $Q^2$, and may also
be used to guide future experiments.

In the present work, we discuss mainly estimates from  
the covariant spectator quark model~\cite{NSTAR2017,Nucleon,Omega}. 
The covariant spectator quark model is a model based on constituent quarks 
where the quark electromagnetic structure is parametrized
in order to describe the nucleon electromagnetic structure. 
The wave functions of the baryons are ruled
by the spin-flavor-radial symmetries~\cite{Capstick00}, 
with radial wave functions 
determined phenomenologically with the assistance of empirical data, 
lattice data or estimates of the 
quark core contributions~\cite{NSTAR,NSTAR2017}. 
One can then use parametrizations of a few resonances $N^\ast$ 
to make predictions for other states based on the symmetries. 
The model is covariant by construction and 
can then be used in applications at large $Q^2$.
In some cases the model can be extended with the inclusion of effective 
descriptions of the meson cloud effects, which can have significant contributions 
at small $Q^2$~\cite{NDelta,NDeltaD,LatticeD,Delta1600,N1520SL,N1520TL,Jido12,N1535TL-new,Octet2Decuplet,Octet2Decuplet2}.

In the next section we describe the formalism associated 
with the covariant spectator quark model.
In Sect.~\ref{secLargeQ2}, we present estimates for 
$\gamma^\ast N \to N^\ast$ transition form factors at large $Q^2$ for 
the states: $\Delta(1232)$, $N(1440)$, $N(1520)$, $N(1535)$, 
$N(1880)$,  $N(1700)$ and $\Delta(1700)$.
In Sect.~\ref{secLowQ2} we present estimates for the 
$\Delta(1232)$ quadrupole form factors at low $Q^2$.
Our summary and conclusions are presented in Sect.~\ref{secConclusions}.

\section{Covariant spectator quark model}
\label{secFormalism}

The covariant spectator quark model~\cite{Nucleon,NSTAR2017,Omega}
is based on the three main ingredients: 
\begin{enumerate} 
\item
the  baryon wave function (including the nucleon) 
is represented in terms of the spin-flavor structure of 
the individual quarks based on 
the $SU_S(2) \times SU_F(3)$ spin-flavor symmetry, 
rearranged as an active quark and a spectator quark-pair~\cite{Nucleon,Omega,Nucleon2};
\item
within the impulse approximation, 
the three-quark system is reduced to a quark-diquark system,
parametrized by a radial wave function $\psi_B$,
after the integration into the quark-pair degrees of freedom~\cite{Nucleon,Nucleon2,Omega}; 
\item
the electromagnetic structure of the quark is 
parametrized by quark isoscalar/isovector and strange quark 
form factors $f_{i \pm}(Q^2)$ and $f_{i s} (Q^2)$
($i=1$ for Dirac, and $i=2$ for Pauli),
which simulate the substructure associated with the gluons 
and quark-antiquark effects,
represented according to the  vector meson mechanism~\cite{Nucleon,Omega,LatticeD,Lattice}.
\end{enumerate}

When the nucleon wave function ($\Psi_N$) and
the resonance wave function ($\Psi_R$) are both expressed in terms 
of the single quark and quark-pair states, 
the transition current in impulse approximation can 
be written as~\cite{Nucleon,Omega,Nucleon2}
\ba
J^\mu=
3 \sum_{\Gamma} 
\int_k \bar \Psi_R (P_R,k) j_q^\mu \Psi_N(P_N,k),
\label{eqJmu}
\ea  
where $P_R$, $P_N$, and $k$ are  
the resonance, the nucleon, and the diquark momenta, respectively.
In the previous equation 
the index $\Gamma$ labels
the intermediate diquark polarization states,
the factor 3 takes account of the contributions from
the other quark pairs by the symmetry, and the integration
symbol represents the covariant integration over the 
diquark on-mass-shell momentum.
In the study of the inelastic transitions, 
we use the Landau prescription to ensure
the current conservation~\cite{SQTM,SemiRel,N1520SL}.

Using Eq.~(\ref{eqJmu}), we can express 
the transition current in terms of the 
quark electromagnetic form factors $f_{i\pm}$ ($i=1,2$)
and the radial wave functions 
$\psi_N$ and $\psi_R$ for resonances 
with no strange quarks
and determine the transition form factors~\cite{Nucleon,SQTM,N1535SL,N1520SL}.
Taking advantage of the quark form factor structure, based on vector 
meson dominance, the model has been extended to the lattice QCD regime 
(heavy pions and reduced meson cloud)~\cite{Omega,LatticeD,Lattice},
to the nuclear medium~\cite{OctetMedium,OctetMedium2} 
and to the timelike regime ($Q^2 < 0$)~\cite{N1520TL,Timelike1,Timelike2,N1535TL-new}.

The present description of the covariant spectator quark model 
takes into account only the effects associated 
with the valence quark degrees of freedom.
There are however some processes, such as
the meson exchanged between the different quarks
inside the baryon, which cannot be reduced
to processes associated with the dressing of a single quark.
Those processes can be regarded  as a consequence 
of the meson exchanged between the different quarks inside
the baryon, and can be classified as meson cloud corrections 
to the hadronic 
reactions~\cite{N1520SL,N1520TL,Octet2Decuplet,Octet2Decuplet2}.

The study of the role of the meson cloud effects 
on the $\gamma^\ast N \to N^\ast$ transition can be done 
also in the context of the 
dynamical coupled-channel reaction models
(dynamical models)~\cite{Burkert04,Kamalov99,SatoLee01,EBAC,Ronchen13}.
Those models use baryon-meson states to 
describe the photo- and electro-production of mesons by nucleons,
taking into account the meson dressing of propagators and vertices,
and are fitted to the data.
The bare core contribution to the data can be determined by removing the effect 
of the meson-baryon dressing of propagators and vertices~\cite{EBAC}. 
Those estimates of the bare core can be used to test 
the limits of models based on valence quarks
or even to calibrate parameters of the quark models.

The separation of the transition form factors 
into the valence quark and meson cloud contributions is
model dependent~\cite{Hammer04,Meissner07},
and the identification of the bare baryon states 
depends of the calibration of the background~\cite{Burkert04,Ronchen13}. 
One can, however, reduce the impact of the model dependence 
comparing our estimates of the valence quark 
contributions with lattice QCD simulations 
for large pion masses.
Although quenched and unquenched 
lattice QCD simulations include some meson cloud 
effects, those effects are reduced 
for large pion masses~\cite{Alexandrou08,Pascalutsa07}.

To compare our model with the lattice QCD simulations, 
we use our calibration of the radial wave functions 
for the baryon lattice QCD masses and the quark current, $j_q^\mu$, expressed 
in terms of the lattice QCD vector meson masses~\cite{Lattice,LatticeD,Omega}.
The free parameters of the radial wave functions 
and quark current are the same in the 
physical and lattice QCD regimes.
The extrapolation of the results for the physical limit
is obtained when we replace the dependence on 
the lattice QCD masses by the physical masses.

%\clearpage

\begin{figure*}[t]
\centering
%\vspace*{5cm}       % Give the correct figure height in cm
\centerline{\mbox{
\includegraphics[width=2.15in]{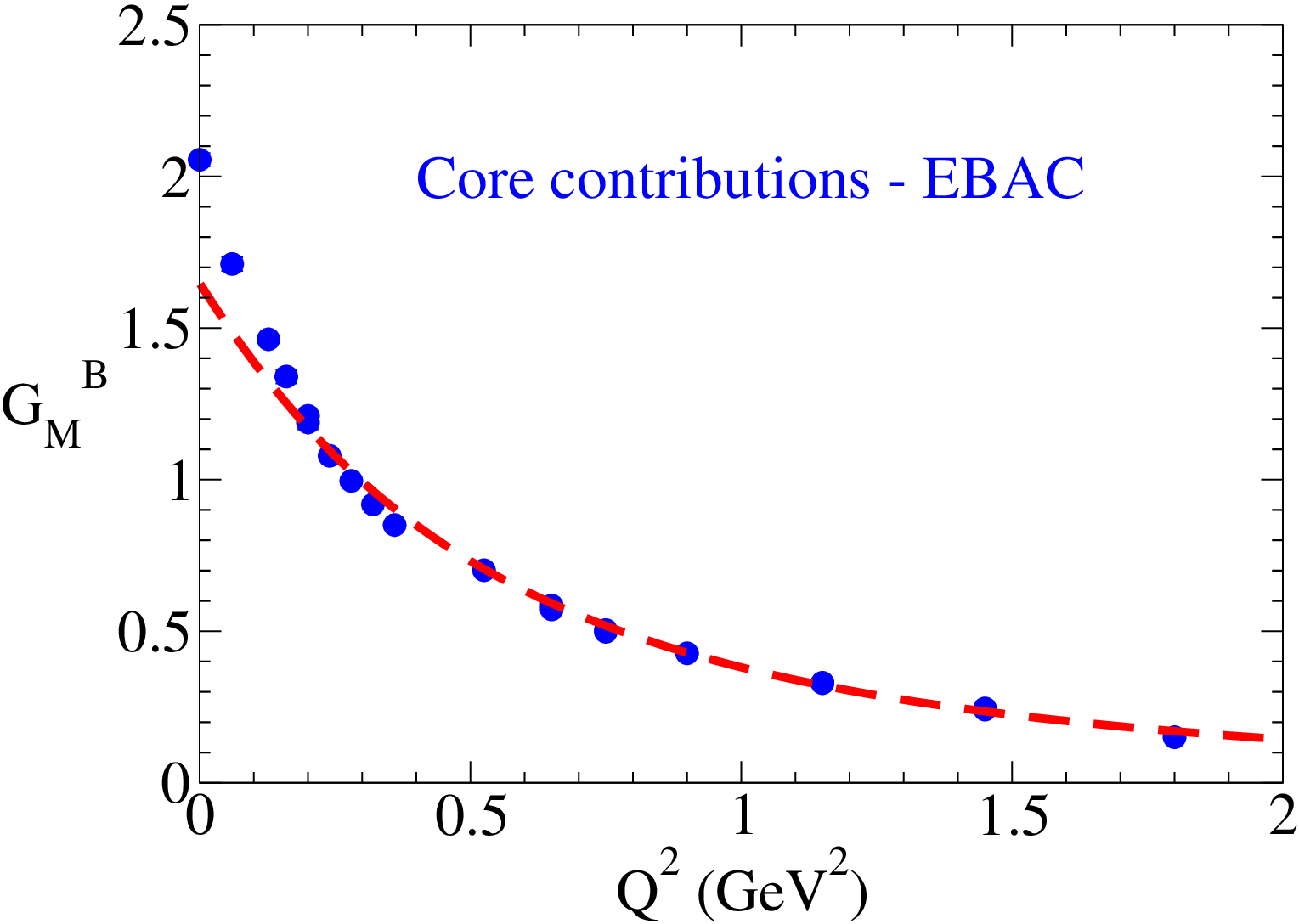}  \hspace{.2cm}
\includegraphics[width=2.1in]{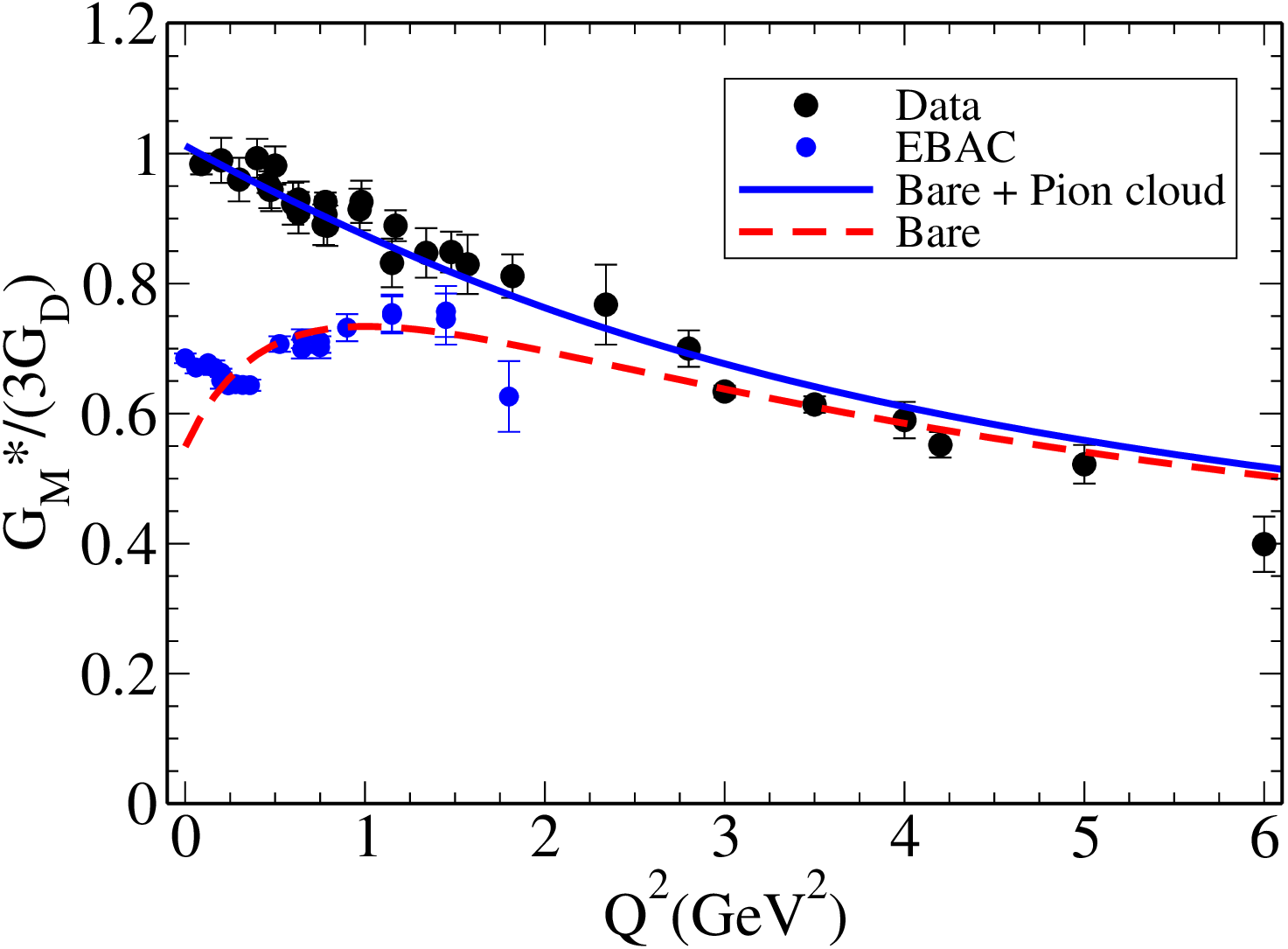}  \hspace{.2cm}
\includegraphics[width=2.05in]{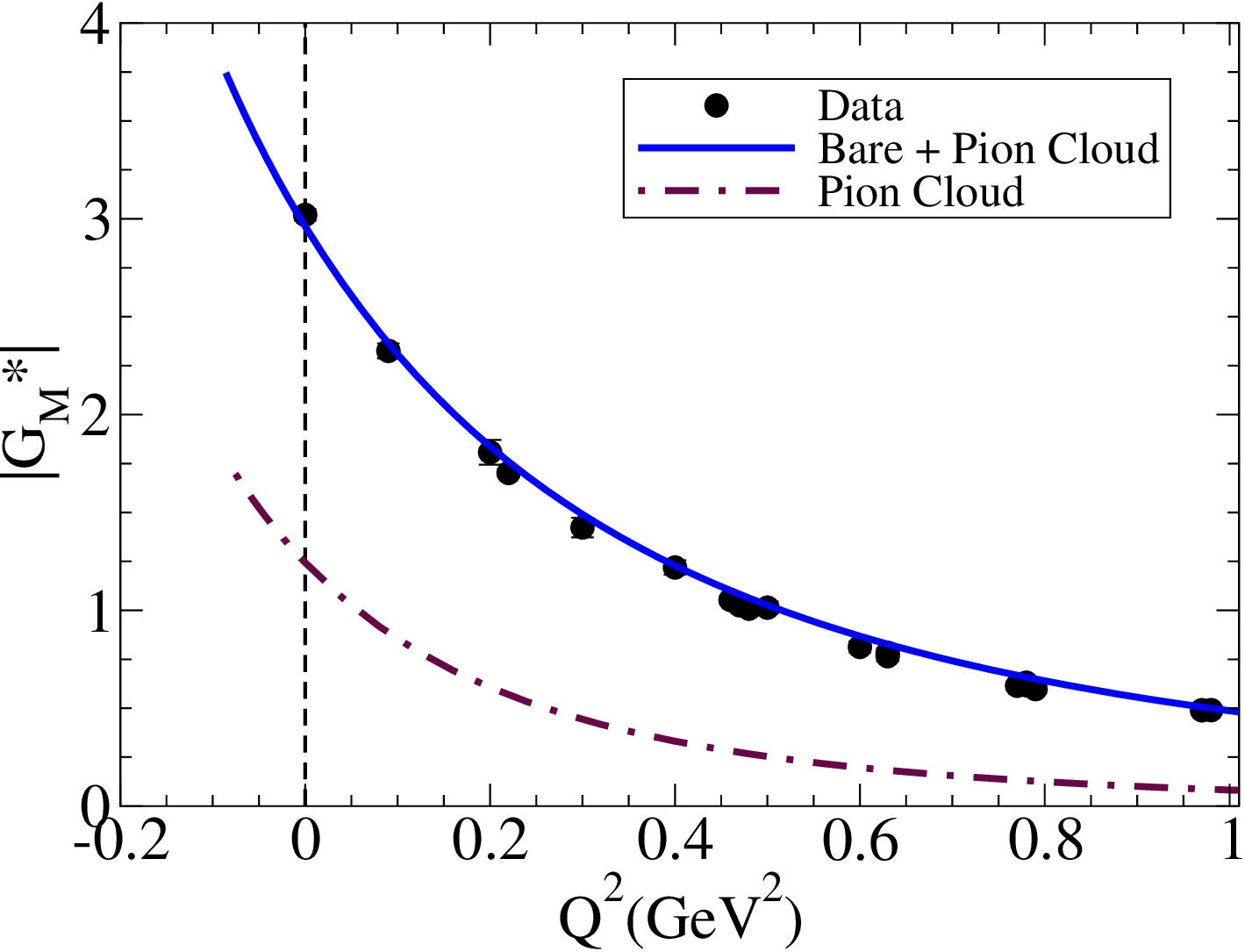}}}
\caption{$\gamma^\ast N \to \Delta(1232)$ magnetic form factor.
{\bf Left:}
Comparison of bare contribution with EBAC estimate from~\cite{EBAC}.
{\bf Center:}
Form factor normalized by 3$G_D$, where $G_D= (1 + Q^2/\Lambda_D^2)^{-2}$,
with $\Lambda_D^2 = 0.71$ GeV$^{2}$.
{\bf Right:}
Re-parametrization of $|G_M^\ast|$ based on expression (\ref{eqGMpi}) 
for the pion cloud 
with $\Lambda_D^2=0.9$ GeV$^2$~\cite{Timelike2}.
Extension to the timelike region.
Database from~\cite{NDelta,NDeltaD}.
}
\label{figDelta}       % Give a unique label
\end{figure*}

\section{Transition form factors at large $Q^2$}
\label{secLargeQ2}

We present next our estimates for 
the transition form factors for 
the $\Delta(1232)$, radial excitations 
of the nucleon and some negative parity states, 
including the states $N(1520)$ and $N(1535)$. 

We use $M_N$ and $M_R$ 
to represent the nucleon and the resonance masses, respectively. 
In the case of the $\Delta(1232)$, we use 
$M_\Delta$, exceptionally.

\subsection{$\gamma^\ast N \to \Delta(1232)$ transition}

The $\gamma^\ast N \to \Delta(1232)$ transition is known to be 
dominated by the magnetic transition~\cite{NSTAR,Jones73},
as a consequence  of a quark spin flip in the nucleon 
to form a $\Delta(1232)$ with spin $\frac{3}{2}$ ($\Delta^+$ or $\Delta^0$).
Quark models, however, underestimate the magnitude 
of the magnetic form factor, $G_M^\ast$, near $Q^2=0$.
This property can be naturally explained in 
the context of the covariant spectator quark model, 
when we consider that the nucleon and the $\Delta(1232)$
systems are mainly composed by $s$-wave quark-diquark states~\cite{NDelta}.
In that case, one concludes, based on the normalization
of the radial wave functions and the   
Cauchy-Schwarz inequality for integrals that 
$G_M^\ast (0) \le 2.07$, well below 
the experimental value 
$G_M^\ast (0) \simeq 3$~\cite{NDelta,NSTAR2017,Octet2Decuplet}.

The missing strength in quark model estimates of $G_M^\ast$
can be understood when we consider the effect of the pion cloud 
dressing in addition to the valence quark effects.
Dynamical models estimate that 
the contribution of the pion cloud effects 
can be about 30-45\% 
of $G_M^\ast$ at low $Q^2$~\cite{Burkert04,Kamalov99,SatoLee01,EBAC}.
The natural conclusion is that quark models 
cannot describe the $\gamma^\ast N \to \Delta(1232)$ completely,
and therefore the $\Delta(1232)$ radial wave function $\psi_\Delta$ 
cannot be extracted directly from the data~\cite{NDelta,LatticeD,NSTAR2017}.

One needs, then to rely on indirect methods to estimate $\psi_\Delta$.
There are two main options: i) use some independent 
estimate of the contribution of the valence quark effects 
to $G_M^\ast$; ii) use lattice QCD simulations of 
the form factors to determine the function $\psi_\Delta$,
assuming that the pion cloud effects 
are small due to the large pion masses.

The first option can be implemented using 
the EBAC/Argonne-Osaka~\cite{EBAC} estimate for 
the bare core contributions, obtained 
when the baryon-meson couplings are turned off~\cite{SatoLee01,EBAC}.
In this limit, we obtain the blue points from Fig.~\ref{figDelta} 
in the left and central panels.
One can see that those points (bare contribution) 
are well described by the fits used 
to  determine  $\psi_\Delta$
(dashed lines in the left and central panels).
At large $Q^2$, we can see in the central 
panel of Fig.~\ref{figDelta}, that the valence quark 
contributions are dominant and  
describe well the data.

% FIG 1 

As mentioned, $\psi_\Delta$, can also be estimated by 
a fit of the model parameters to a lattice QCD data
with \mbox{$m_\pi \simeq $ 400--600 MeV}~\cite{LatticeD,Alexandrou08}.
In this case, the pion cloud effects 
are reduced and one obtains a cleaner estimate 
of the valence quark effects.
Based on the extension of the model to the lattice
QCD regime~\cite{Lattice,LatticeD}, 
we conclude that the parametrization derived from 
the EBAC estimate is also consistent with the lattice QCD data 
from~\cite{Alexandrou08}, as shown in~\cite{LatticeD}.
The main conclusion is then that $\psi_\Delta$ can be 
determined either by lattice QCD data or 
by the EBAC estimate of the bare core.

To describe the $G_M^\ast$ data, we need to include some estimate 
of the pion cloud effects ($G_M^\pi$). 
Since our framework is based on valence quarks, 
we choose to simulate the pion cloud effects by a simple 
parametrization: 
$G_M^\pi = 3 \lambda_\pi G_D \left( \frac{\Lambda_\pi^2}{\Lambda_\pi^2 + Q^2} \right)^2$,
where $\lambda_\pi \simeq 0.448$ and $\Lambda_\pi^2 \simeq 1.53$ GeV$^2$~\cite{NDeltaD}.
The final result for $G_M^\ast$ is represented by the solid line 
in the central and right panels of Fig.~\ref{figDelta}.
In the more recent applications of the model, 
which include the $\gamma^\ast N \to \Delta(1232)$ transition 
in the timelike region~\cite{Timelike1,Timelike2}, 
we consider an alternative parametrization based 
on the connection 
with the microscopic pion cloud mechanisms~\cite{Octet2Decuplet,Timelike2}.
We consider in particular, the parametrization 
\ba
G_M^\pi &=&
\frac{3}{2}\lambda_\pi 
F_\pi (q^2) \left( \frac{\Lambda_\pi^2}{\Lambda_\pi^2-q^2}\right)^2 +
\nonumber \\
& &  \frac{3}{2}\lambda_\pi 
\frac{\Lambda_D^4}{(\Lambda_D^2 -q^2)^2 + 
\Lambda_D^2 [\Gamma_D(q^2)]^2 }, 
\label{eqGMpi}
\ea
where the first term is associated with the photon-pion coupling,
proportional to the pion electromagnetic form factor $F_\pi (q^2)$,
and the second term is associated with the photon coupling with 
the intermediate baryon states.
This last coupling is simulated by an effective dipole function $G_D$ 
with $\Lambda_D^2 = 0.90$ GeV$^2$, and include an
effective width $\Gamma_D$ in order to avoid singularities 
in the region \mbox{$q^2=-Q^2  > 0$.} 
The new parametrization is presented in the right 
panel of Fig.~\ref{figDelta} (dashed-dotted line). 
The advantage of the last parametrization of $G_M^\pi$ 
is that the model can be naturally extended to the timelike region,
above $Q^2= -(M_\Delta -M_N)^2$.
These estimates are useful for the study of 
the $\Delta$ Dalitz decay ($\Delta \to e^+ e^- N$) 
measured at HADES.
The comparison of our calculations with 
the HADES $\Delta(1232)$ Dalitz decay cross-sections 
can be found in~\cite{HADES17}.
Those results are also discussed in 
the presentations of 
B.~Ramstain and P.~Salabura~\cite{Beatrice-talk,Salabura-talk}.  

% D1, D3 states:  a ~ 0.085,  b ~ 0.085  
%                 a^2 ~ b^2 ~ 0.72%

The present discussion of the $\gamma^\ast N \to \Delta(1232)$ transition 
is simplified because we ignore the contributions 
of the $\Delta(1232)$ $d$-wave states~\cite{NDeltaD,DeltaDFF,DeltaShape,LatticeD}.
The procedure is justified because the admixture terms 
are of the order of 0.7\% for both states.
Those states are, however, important for the sub-leading 
transition form factors, the electric ($G_E^\ast$) 
and Coulomb ($G_C^\ast$) quadrupole form factors.
Those form factors are described in the Sect.~\ref{secLowQ2}, at low $Q^2$.

\begin{figure*}
\centering
%\vspace*{5cm}       % Give the correct figure height in cm
%\includegraphics[width=2.5in]{figs_NSTAR2019/AmpA12_Roper2}  \hspace{.1cm}
%\includegraphics[width=2.5in]{figs_NSTAR2019/AmpS12_Roper2}
\includegraphics[width=7cm]{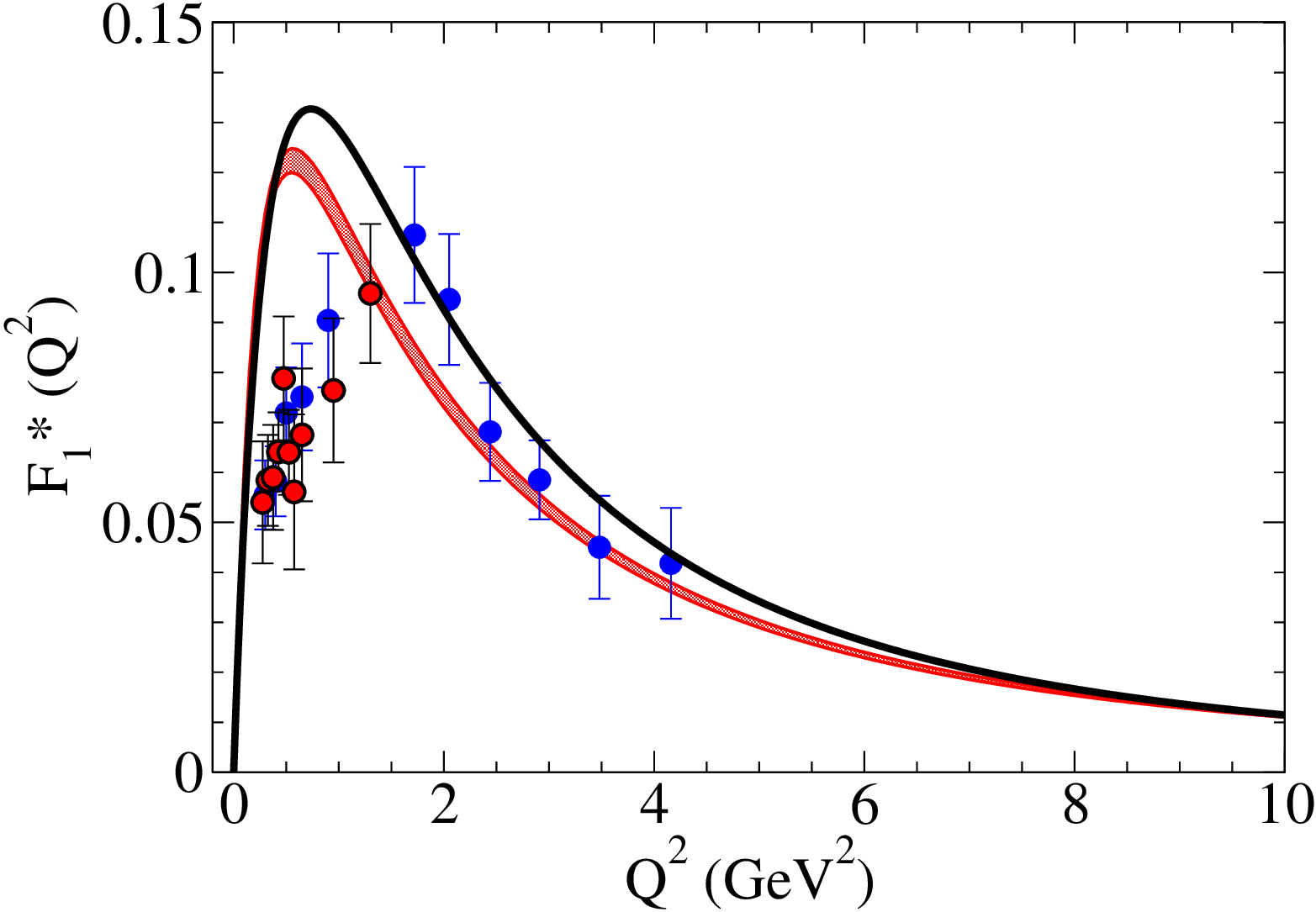}  \hspace{.8cm}
\includegraphics[width=7cm]{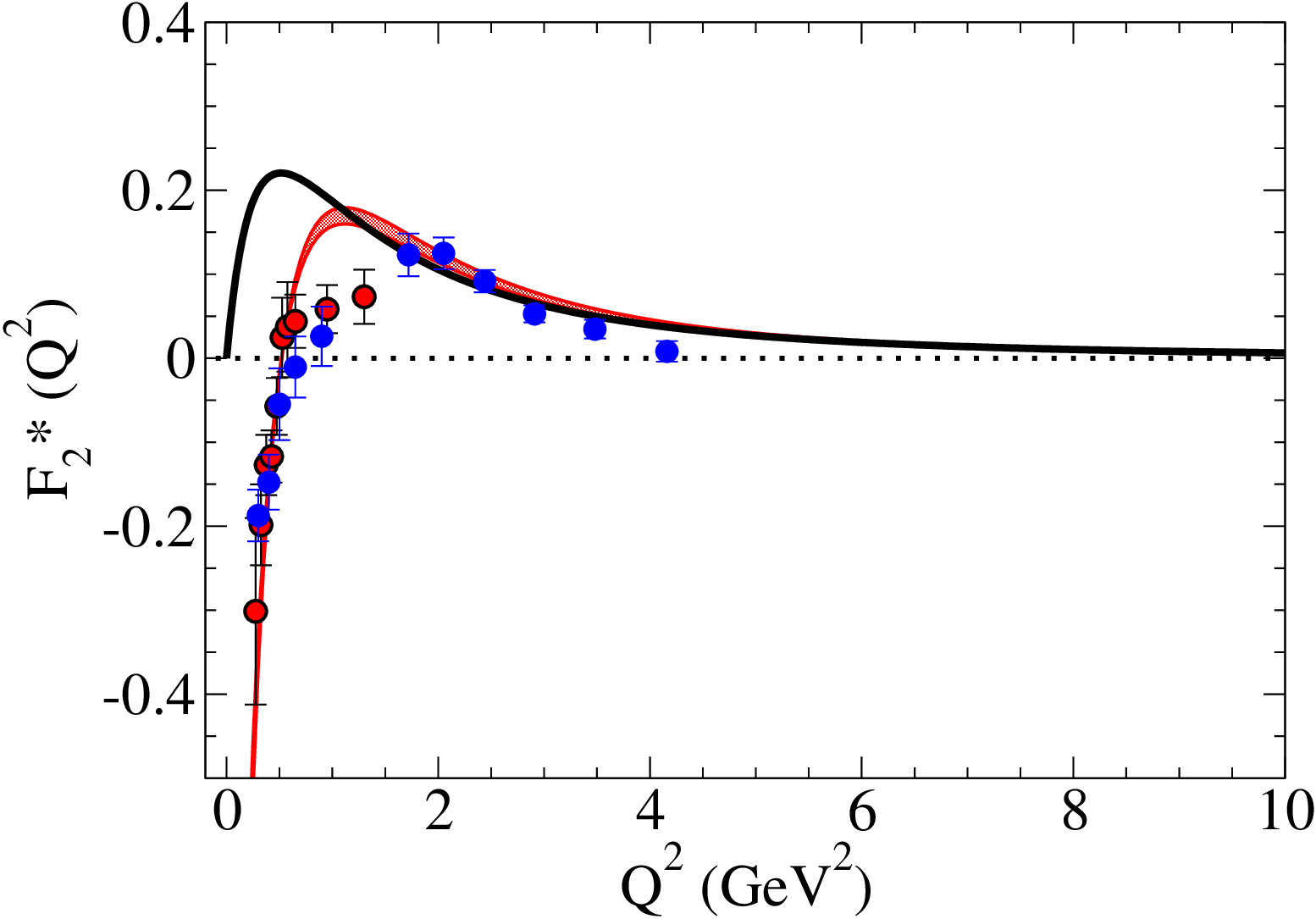}
\caption{$\gamma^\ast N \to N(1440)$ transition form factors.
The solid line represent the estimate of the 
covariant quark model~\cite{Roper}.
The red band represent the estimated from AdS/QCD 
taking into account the uncertainty of the couplings~\cite{Roper-AdS}.
Data from CLAS~\cite{CLAS09,CLAS12,CLAS16}. 
}
\label{figRoper}       % Give a unique label
\end{figure*}

\begin{figure*}
\centering
%\vspace*{5cm}       % Give the correct figure height in cm
%\includegraphics[width=2.4in]{figs_NSTAR2019/AmpA12_Roper2}  \hspace{.1cm}
%\includegraphics[width=2.4in]{figs_NSTAR2019/AmpS12_Roper2}
\includegraphics[width=7cm,clip]{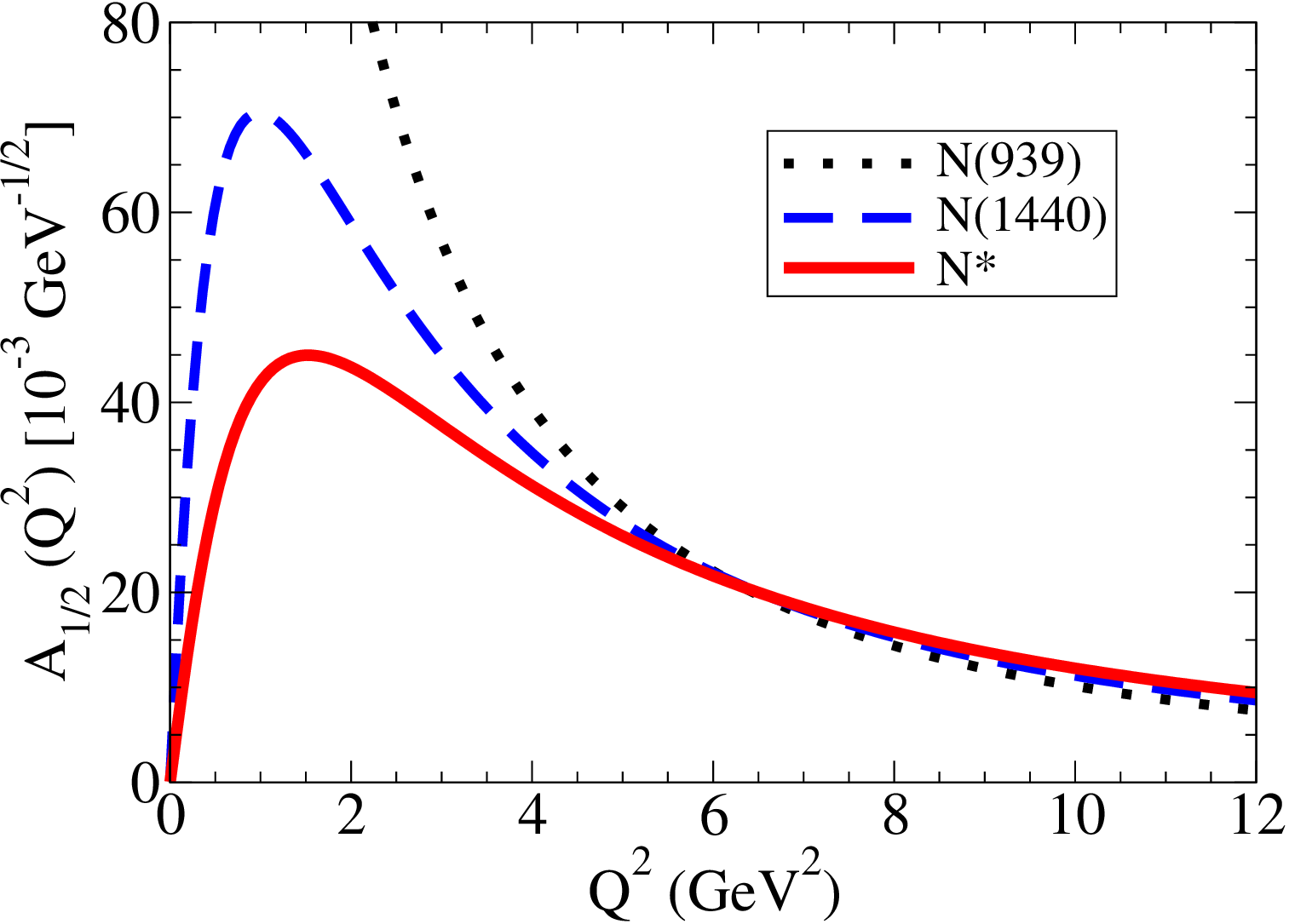}
\hspace{.8cm}
\includegraphics[width=7cm,clip]{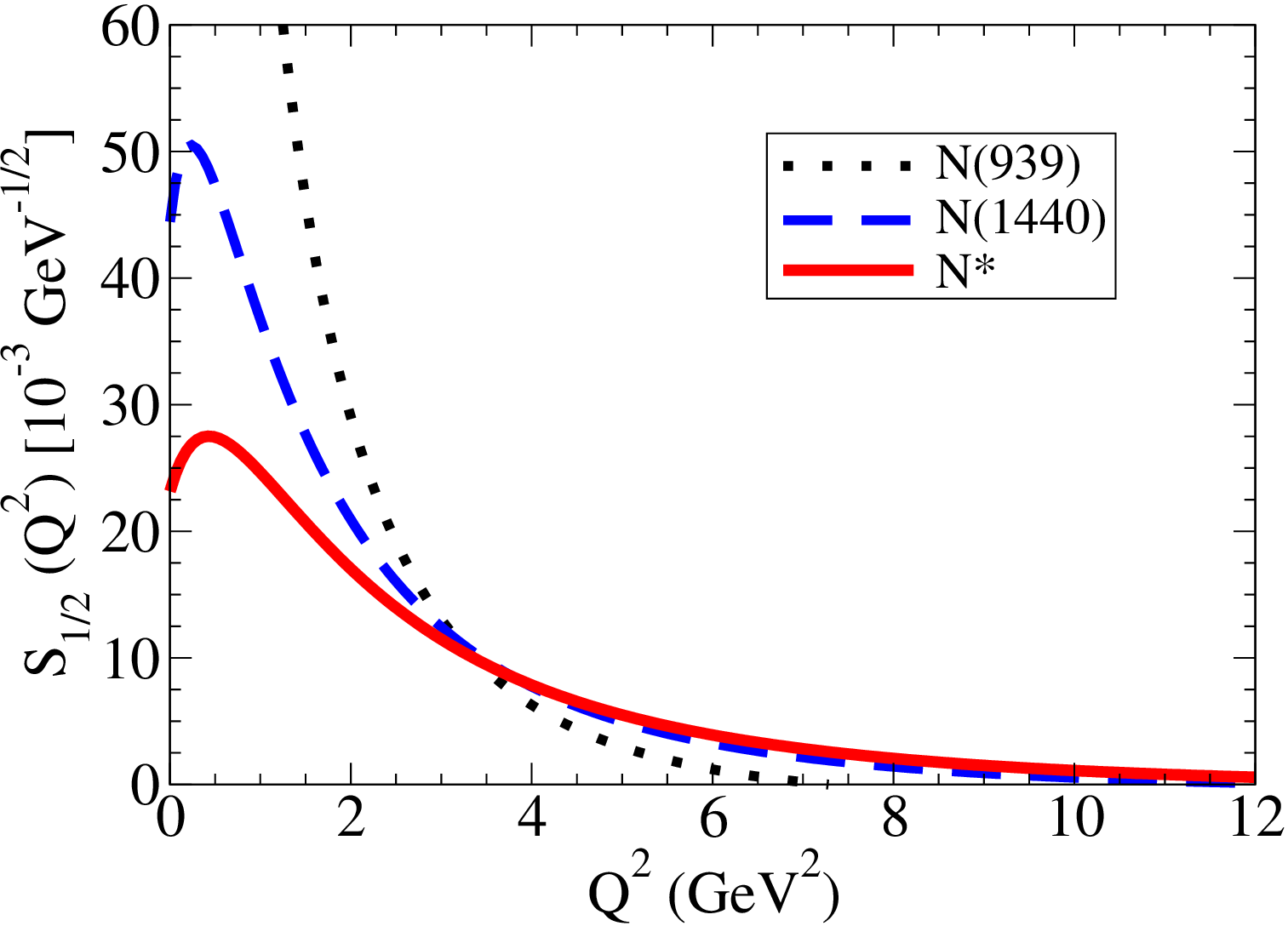}
\caption{$\gamma^\ast N \to N^\ast$ transition amplitudes
for the first and the second ($N^\ast$) radial excitations
of the nucleon. 
$N(1930)$ represent the nucleon's equivalent amplitudes 
(description in the main text).}
\label{figRoper2}       % Give a unique label
\end{figure*}

\subsection{Radial excitations}

The covariant spectator quark model can also be 
applied to the radial excitations of the nucleon.
The first radial excitation, $N(1440)$ or Roper,
can be defined as a system which differs from 
the nucleon only by the radial structure, and has a zero 
in the radial wave function~\cite{Roper,Roper2}.
The $N(1440)$ state can then be determined 
defining a radial wave function $\psi_R$ 
which is orthogonal to the nucleon and has a single node.
The transition form factors are then calculated without any  
adjustable parameters.
The results (solid line) are presented in Fig.~\ref{figRoper}
in comparison with the CLAS data~\cite{CLAS09,CLAS12,CLAS16}.

From the graphs, we conclude that the model describe
well the data above 1.5 GeV$^2$ for both form factors.
Those results are an indication that $N(1440)$ 
is in fact the first radial excitation of the nucleon.
The disagreement at low $Q^2$ can be a consequence 
of not taking into account possible meson cloud effects~\cite{Roper,Roper2}.

The $\gamma^\ast N \to N(1440)$ form factors 
have been also estimated within the AdS/QCD framework~\cite{Roper-AdS,Roper-AdS2}.
In the leading twist approximation, 
where the baryons correspond to three-quark systems, 
the nucleon and the Roper transition form factors depend 
on three independent couplings, which can be fixed 
by the nucleon elastic form factor data,
within certain limits.
The results for the Roper are then determined without 
any additional fit. 
Those are also represented in Fig.~\ref{figRoper} 
by the red band between the upper and lower limits.
The results of the Pauli form factor $F_2^\ast$ are in excellent
agreement with the data.
Overall, it is interesting to observe 
that the two estimates converge to the same result 
for large values of $Q^2$, which can be seen 
as a consequence of the dominance of the valence quark degrees 
of freedom at large $Q^2$.

The covariant spectator quark model can also be used to 
estimate the second radial excitation of the nucleon.
In a previous work~\cite{Roper2}, we assumed tentatively that 
the state could be the $N(1710)\frac{1}{2}^+$, but the assumption was discarded 
by recent measurements from CLAS~\cite{Park15}.
The next candidate is then the state $N(1880)\frac{1}{2}^+$.
In this case, we compare the results 
directly to the helicity amplitudes 
$A_{1/2} \propto (F_{1p}^\ast +  F_{2p}^\ast)$
and $S_{1/2} \propto (F_{1p}^\ast - \tau  F_{2p}^\ast)$,
where $\tau= \frac{Q^2}{(M_N + M_R)^2}$.
The results for the Roper and the next radial excitation ($N^\ast$)
are presented in Fig.~\ref{figRoper2}.
In the graphs, we include also the equivalent amplitudes 
of the nucleon $A_{1/2} \propto G_M$ and  $S_{1/2} \propto G_E$
(the conversion factors are calculated using 
the mass of the Roper)~\cite{Roper2}.
Note that, the helicity amplitude representation
is particularly useful, because 
it emphasizes the similarity of the amplitudes associated 
with the different states at large $Q^2$, particularly to $A_{1/2}$.
The corollary of this result is that, 
one can use the experimental results of
the magnetic form factor of the proton,  $G_{Mp}$,
to predict the amplitude $A_{1/2}$ for the Roper 
and $N(1880)\frac{1}{2}^+$ states in the range $Q^2=10$--30 GeV$^2$.
Hopefully our predictions for $A_{1/2} \propto G_{Mp}$ 
can be tested in a near future. 

Similarly to the case of the nucleon, 
we can use the formalism to estimate 
the first radial excitation of the $\Delta(1232)$,
the state $\Delta(1600)$, based on our model for the $\Delta(1232)$.
Our estimates are valid for large $Q^2$ and differ 
from the present measurements at $Q^2=0$,
although the estimates approach the experimental data 
when we take into account the meson cloud dressing 
based on a simplified model derived from the 
$\gamma^\ast N \to \Delta(1232)$ pion cloud parametrization~\cite{Delta1600}.
Our estimates for the $\gamma^\ast N \to \Delta(1600)$ 
form factors may be tested in a near future,
since data are under analysis and results are expected soon~\cite{Ralf-talk}.

\subsection{$\gamma^\ast N \to N(1520)$ 
and $\gamma^\ast N \to N(1535)$ transitions}

\begin{figure*}[t]
%\vspace{.6cm}
\centerline{\mbox{
\includegraphics[width=2.1in]{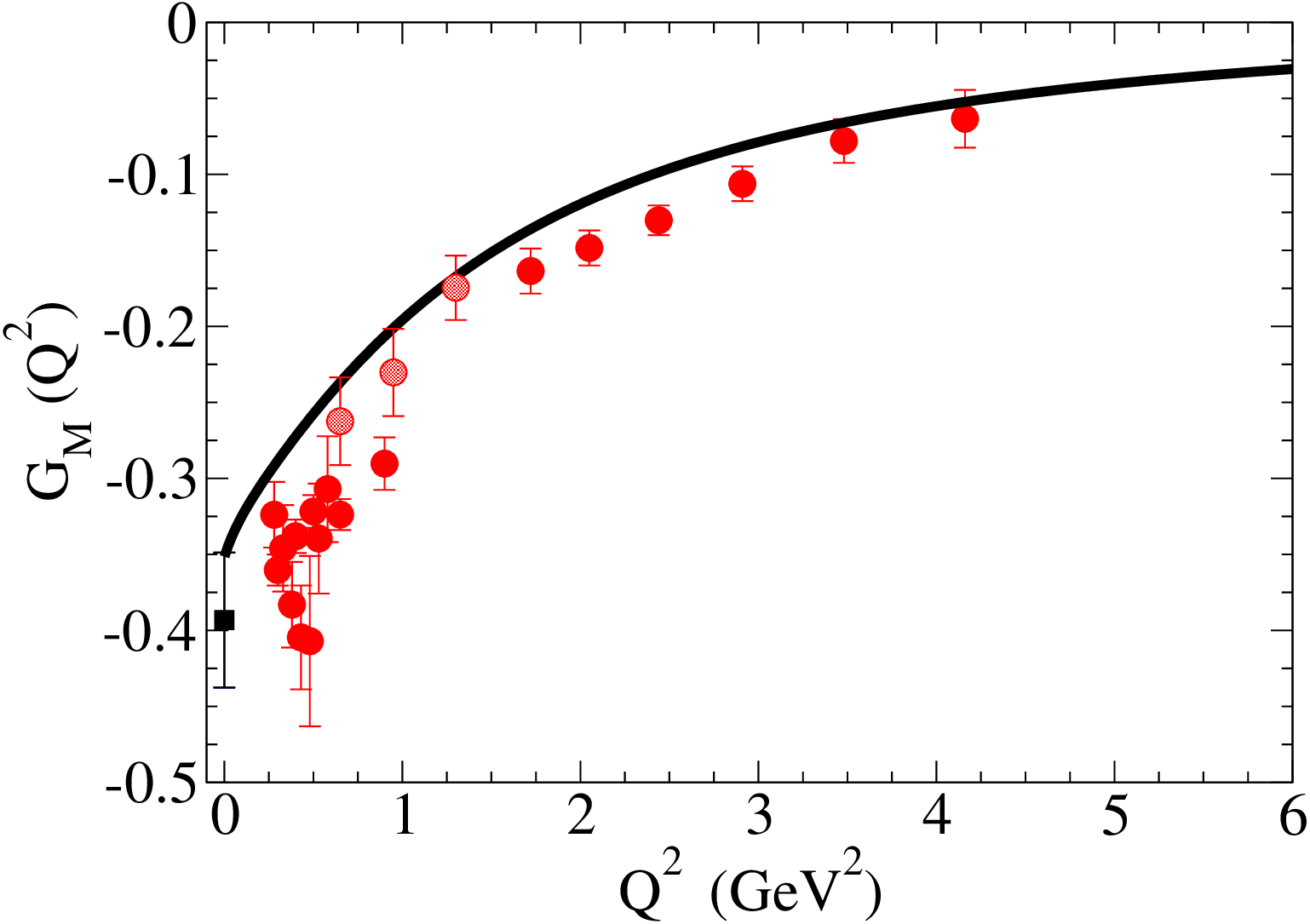}  \hspace{.2cm}
\includegraphics[width=2.1in]{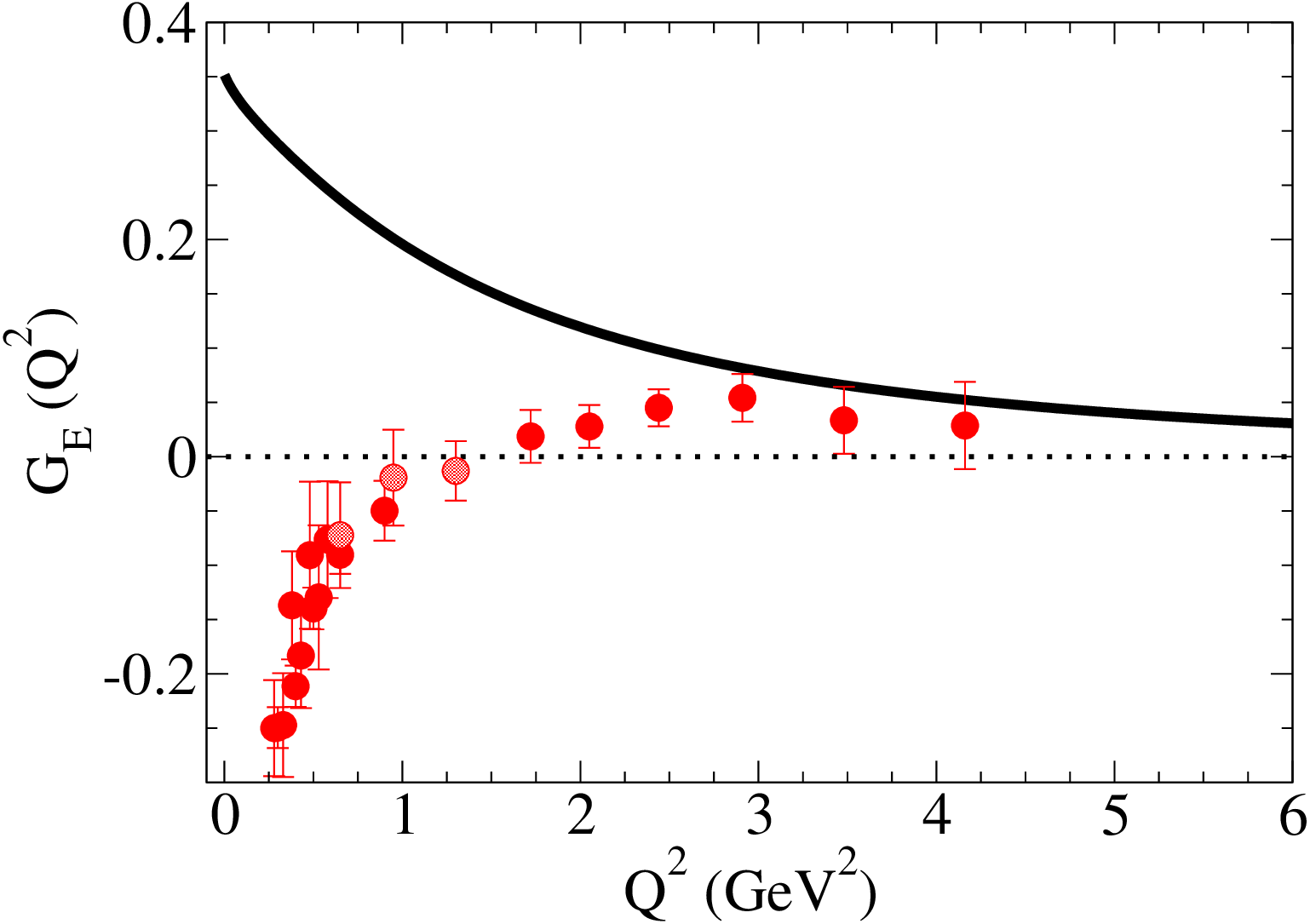}  \hspace{.2cm}
\includegraphics[width=2.1in]{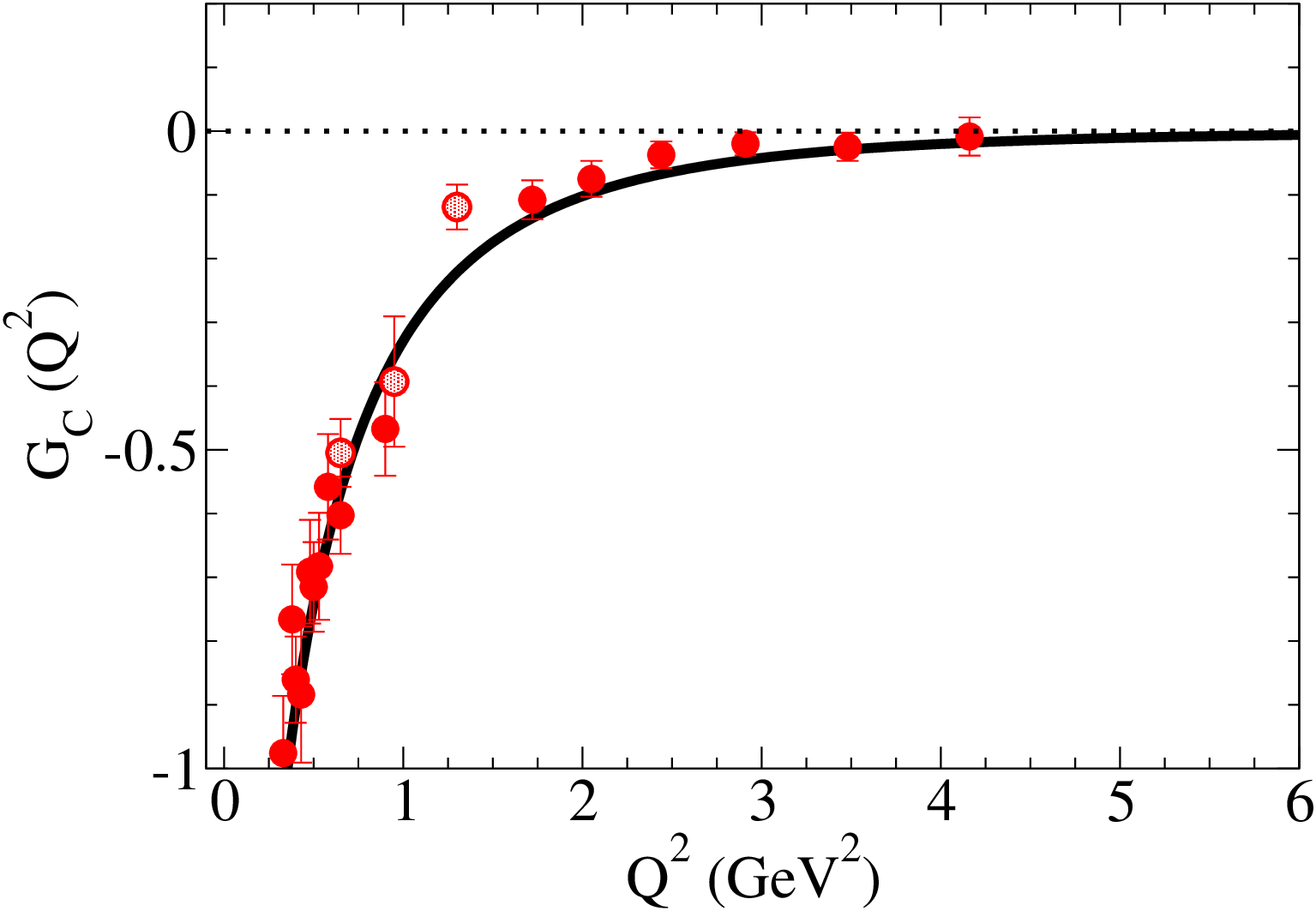}}}
\caption{\footnotesize
$\gamma^\ast N \to N(1520)$ transition form factors.
The solid line is the result of the semirelativistic approximation.
Data from JLab (circles)~\cite{MokeevDatabase,CLAS09,CLAS16}
and PDG (square)~\cite{MokeevDatabase}.}
%\vspace{-1cm}
\label{figNegativeP2}
\end{figure*}

\begin{figure*}[t]
%\vspace{.6cm}
\centerline{\mbox{
\includegraphics[width=7cm]{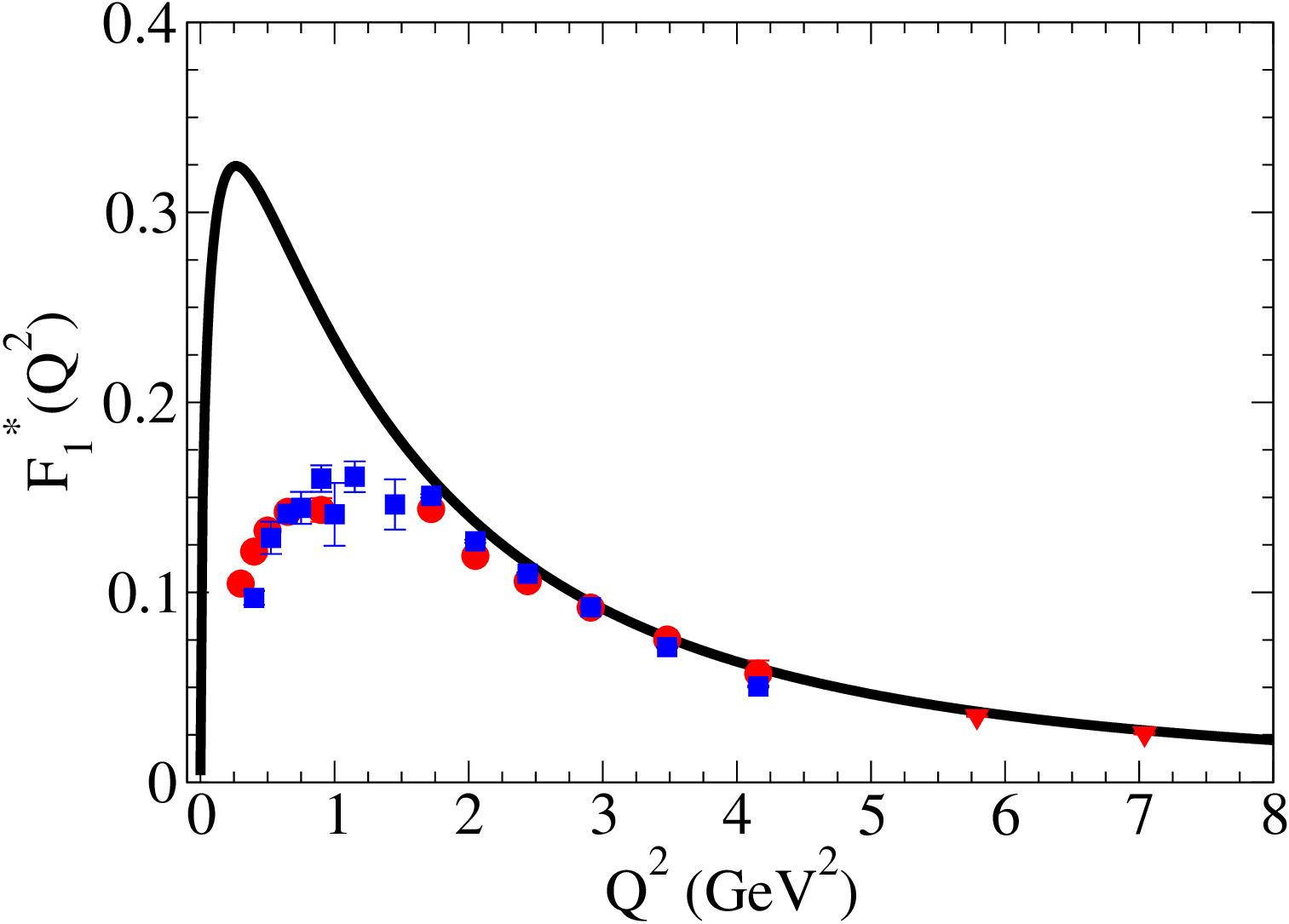}   \hspace{1cm}
\includegraphics[width=7cm]{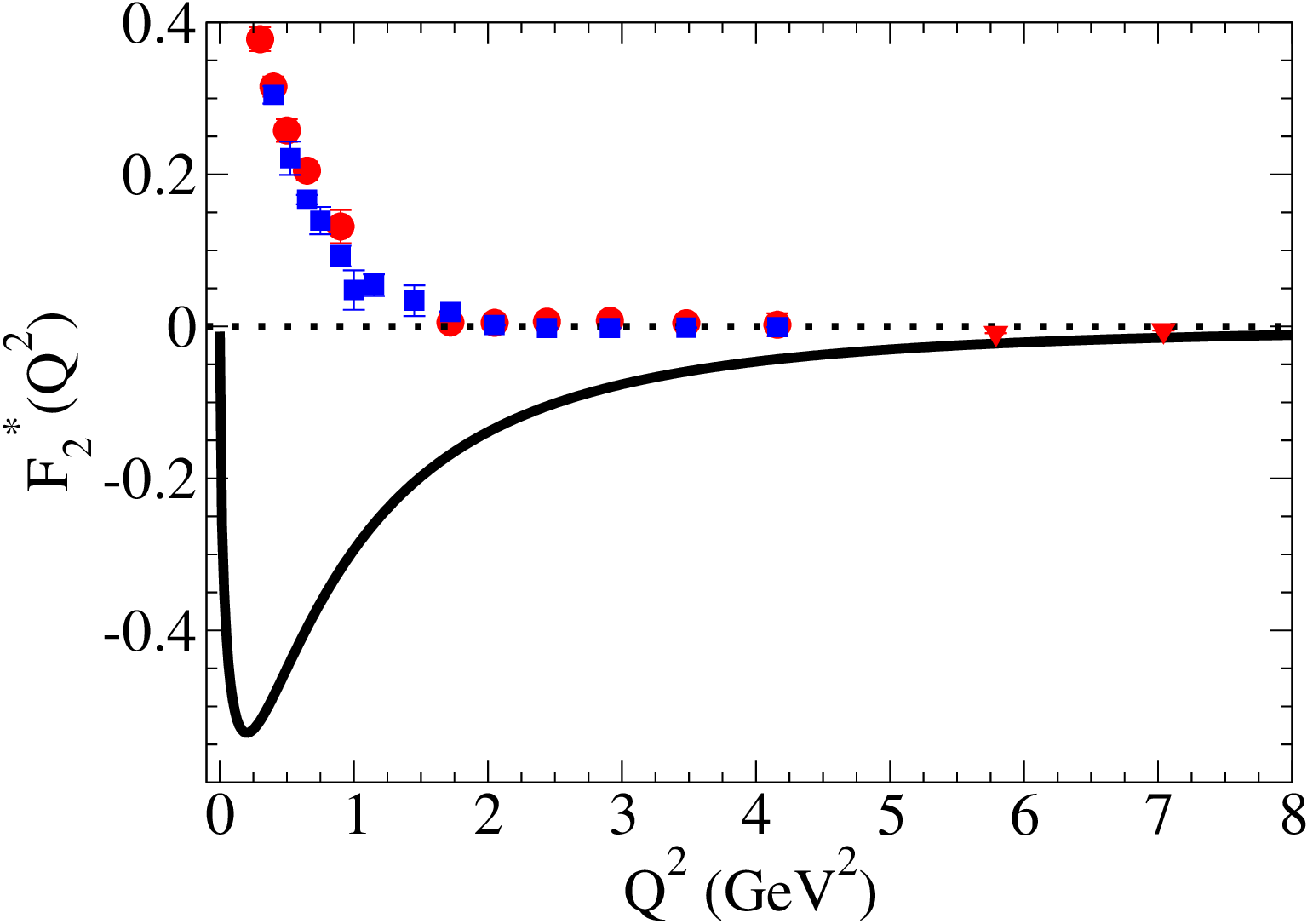}}}
\caption{\footnotesize
$\gamma^\ast N \to N(1535)$ Dirac transition form factor.
Data from JLab (circles and triangles)~\cite{MokeevDatabase,CLAS09,Dalton09} 
and MAID (squares)~\cite{Drechsel07}.
}
%\vspace{-1cm}
\label{figNegativeP}
\end{figure*}

The $N(1520)$ and $N(1535)$ are negative parity states.
Applications of the covariant spectator quark model 
to the $\gamma^\ast N \to N(1535)$ and $\gamma^\ast N \to N(1520)$
transitions are discussed in Refs.~\cite{N1535SL,N1520SL,N1520TL}
based on different prescriptions to the 
radial wave functions.

In particular the  $\gamma^\ast N \to N(1520)$ transition 
was extended to the timelike region~\cite{N1520TL} 
and to the study of the HADES $N(1520)$ Dalitz decays.
Some preliminary results are discussed 
in the B.~Ramstein presentation~\cite{Beatrice-talk}.

Recently the 
$\gamma^\ast N \to N(1535)$ and $\gamma^\ast N \to N(1520)$ transitions
have been studied within the semirelativistic approximation.
The semirelativistic approximation is characterized 
by the following properties~\cite{SemiRel}:
i) the mass difference ($M_R$ and $M_N$) 
is neglected in a first approximation;
ii) the radial wave function of the resonance 
$\psi_R$ has the same form as the nucleon radial 
wave function.

Under these conditions the orthogonality 
between the states is naturally ensured,
and the transition form factors are calculated 
without the introduction of any additional parameters~\cite{SemiRel}.
Those estimates are therefore true predictions.
All model parameters are determined by 
the original parametrization of the nucleon system.
The semirelativistic estimates for the 
$\gamma^\ast N \to N(1520)$ and $\gamma^\ast N \to N(1535)$ 
form factors are presented in Figs.~\ref{figNegativeP2}
and \ref{figNegativeP}, respectively.

In Fig.~\ref{figNegativeP2} we compare 
our estimates for the  $\gamma^\ast N \to N(1520)$ 
transition form factors: $G_M$ (magnetic), $G_E$ (electric) and $G_C$ (Coulomb)
with the available data.
One can notice that one obtain a good description 
of the data for $G_M$ and $G_C$, particularly for $Q^2 > 1$ GeV$^2$.
As for $G_E$ the estimate fails for $Q^2 < 2.5$ GeV$^2$,
which seems to be a consequence of the omission 
of the meson cloud effects.
In the context of the covariant spectator quark model
the result for $G_E$ is a consequence of the estimate $A_{3/2} \equiv 0$,
indicating that the amplitude  $A_{3/2}$ is dominated 
by meson cloud effects~\cite{N1520SL,N1520TL}.

In Fig.~\ref{figNegativeP}, we compare 
the estimates of the Dirac ($F_1^\ast$) and Pauli ($F_2^\ast$)
$\gamma^\ast N \to N(1535)$
with the available data.
From the graph for $F_1^\ast$,  we conclude that the model 
describe well the data above 1.5 GeV$^2$.
From the graph for $F_2^\ast$, we can notice 
that the model fails completely the description of the data (wrong sign).
These two discrepancies may be an indication 
of the impact of the meson cloud effects, not included in 
the present calculations.
There are evidences that the inclusion of 
meson cloud effects can in fact improve 
the description of the 
data~\cite{Jido08,Jido12,N1535TL-new,N1535SL,N1535-S12,SQTM}.

\begin{figure}[b] %[b]
\centering
\includegraphics[width=8cm,clip]{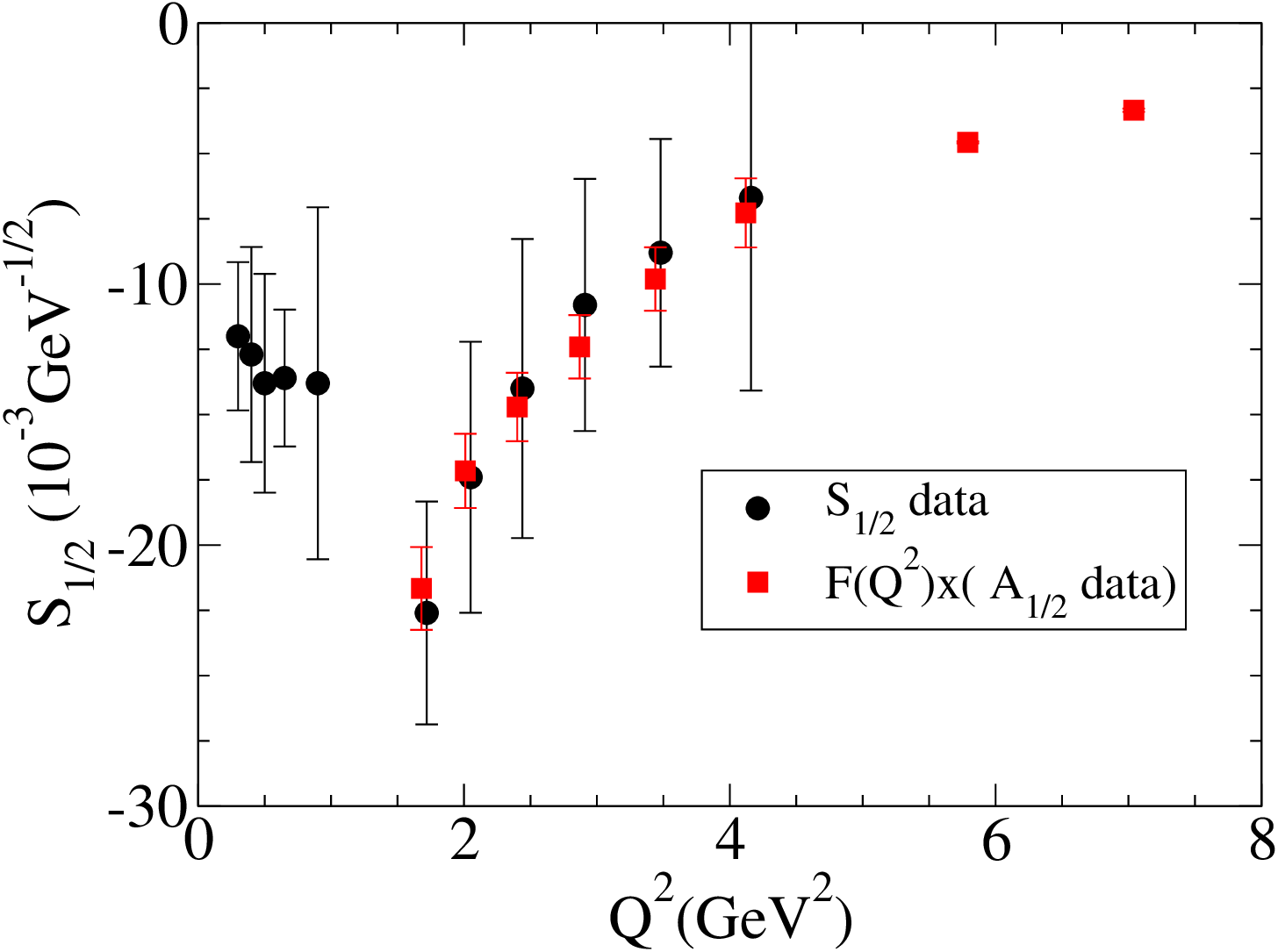}
\caption{$\gamma^\ast N \to N(1535)$ amplitude $S_{1/2}$.
Data compared with estimate from Eq.~(\ref{eqS12}). 
Data from CLAS~\cite{CLAS09,Dalton09}.}
\label{figN1535}       % Give a unique label
\end{figure}

%\newpage

\vspace{.2cm}

The results for $F_2^\ast$ in particular raises the possibility that
there are cancellations between the valence quark effects 
(estimated by our model) and the meson cloud effects 
(not included in our estimate).
In that case, we can explain the 
experimental result $F_2^\ast \approx 0$ for $Q^2 > 1.5$ GeV$^2$.
The implication of the previous result is that 
the $\gamma^\ast N \to N(1535)$ amplitudes are 
related by~\cite{N1535-S12}
\ba
S_{1/2} = - \frac{\sqrt{1 + \tau}}{\sqrt{2}} 
\frac{M_R^2 -M_N^2}{2 M_R Q} A_{1/2},
\label{eqS12}
\ea
for $Q^2 > 1.5$ GeV$^2$, where $\tau = \frac{Q^2}{(M_R + M_N)^2}$ 
and $Q=\sqrt{Q^2}$.

\begin{figure*}[t]
\centering
%\vspace*{5cm}       % Give the correct figure height in cm
%\includegraphics[width=2.5in]{figs_NSTAR2019/AmpA12_Roper2}  \hspace{.1cm}
%\includegraphics[width=2.5in]{figs_NSTAR2019/AmpS12_Roper2}
\includegraphics[width=7cm]{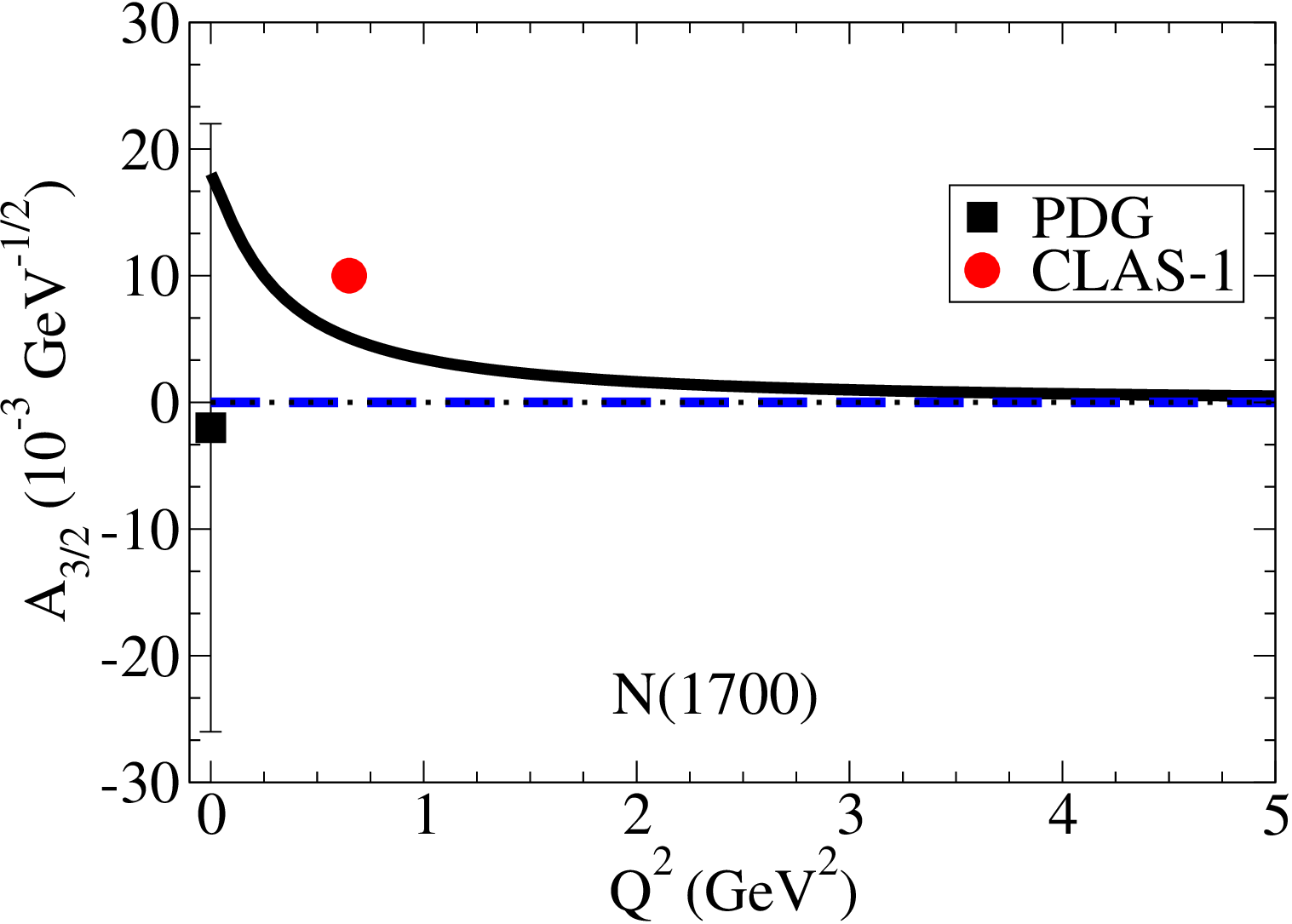}  \hspace{1cm}
\includegraphics[width=7cm]{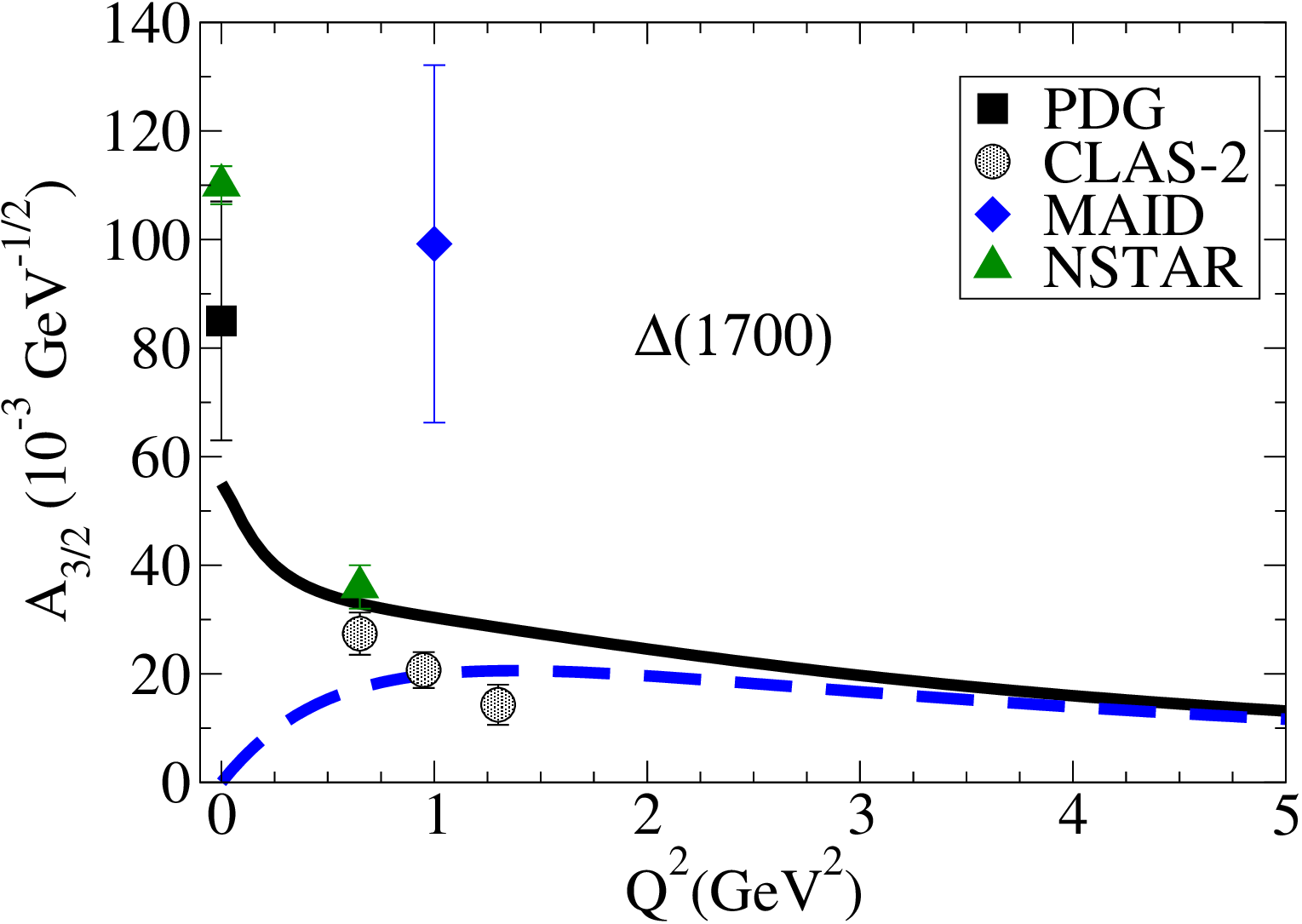}
\caption{Examples of estimates of amplitudes associated 
with negative parity states based on the single quark transition model~\cite{SQTM}.
The solid line represent a model where the meson cloud effects are taken 
into account. Check~\cite{SQTM} for the description of the data.}
\label{figSQTM}       % Give a unique label
\end{figure*}

The relation (\ref{eqS12}) is tested in Fig.~\ref{figN1535}
based on the data for the amplitude $A_{1/2}$.
The last two point present a prediction for the $S_{1/2}$ data.
New large $Q^2$ data are needed in order to test 
the relation (\ref{eqS12}) for higher values of $Q^2$.

The results for the  $\gamma^\ast N \to N(1535)$ transition 
motivated the use of the valence quark model to 
calibrate the meson cloud contributions and 
the extension of the results to the 
timelike region and the $N(1535)$ Dalitz decay~\cite{N1535TL-new}.

Overall, we can conclude that the semirelativistic approximation 
provides a good description of the  $\gamma^\ast N \to N(1520)$ 
and $\gamma^\ast N \to N(1535)$ 
transition form factors for $Q^2 >$ 1.5 GeV$^2$.
The exceptions are the Pauli form factor for $N(1535)$
and the electric form factor for $N(1520)$,
the two cases for which there are indications 
that meson cloud effects may be more significant.
We emphasize that all the calculations 
include no free parameters, apart the calibration
of the nucleon system.
A more detailed discussion of the results for the  $\gamma^\ast N \to N(1535)$ 
and $\gamma^\ast N \to N(1520)$ transitions can be found 
in~\mbox{\cite{N1535SL,N1520SL,SQTM,N1535-S12,Jido12,Jido08,N1535TL-new}.}

%\newpage

\subsection{Single Quark Transition Model}

The covariant spectator quark model can also be used 
in combination with single quark transition model (SQTM)~\cite{SQTM,Burkert03},
since the two models are based on similar principles:
electromagnetic interaction in impulse approximation 
and wave functions determined by the $SU(6) \otimes O(3)$ 
structure~\cite{Capstick00}.

Of particular interest is the $[70,1^-]$ supermultiplet, 
whose transverse amplitudes, $A_{1/2}$ and $A_{3/2}$, 
can be determined by three independent functions of $Q^2$:
$A$, $B$ and $C$, according with the SQTM.
In these circumstances, one can determine
the transverse amplitudes of the states
$N(1650)\frac{1}{2}^-$, $\Delta(1620)\frac{1}{2}^-$,
$N(1700)\frac{3}{2}^-$ and $\Delta(1700)\frac{3}{2}^-$,
once the functions $A$, $B$ and $C$ are known.
These three functions can be calculated using 
the covariant spectator quark models for 
the $N(1520)\frac{3}{2}^-$ and  $N(1535)\frac{1}{2}^-$
transverse amplitudes~\cite{SQTM,NSTAR2017,N1535SL,N1520SL}.

The estimates for the  $N(1650)\frac{1}{2}^-$
and $\Delta(1620)\frac{1}{2}^-$ amplitudes are 
consistent with available data~\cite{Drechsel07,MokeevDatabase} 
for large $Q^2$~\cite{SQTM}.
For the remaining cases, there are no available data for $Q^2 > 2$ GeV$^2$.
As an example, we present in Fig.~\ref{figSQTM}, the results for 
the amplitude $A_{3/2}$ associated with the states $N(1700)$ and $\Delta(1700)$.
Note in both cases the lack of data for $Q^2> $ 1.5 GeV$^2$.
Our estimates can be tested, then only when new large $Q^2$ data 
from the JLab-12 GeV upgrade became available~\cite{NSTAR}.

%\clearpage

\section{Transition form factors at low $Q^ 2$}
\label{secLowQ2}

\begin{figure*}[t]
\centering
%\vspace*{5cm}       % Give the correct figure height in cm
\includegraphics[width=7cm,clip]{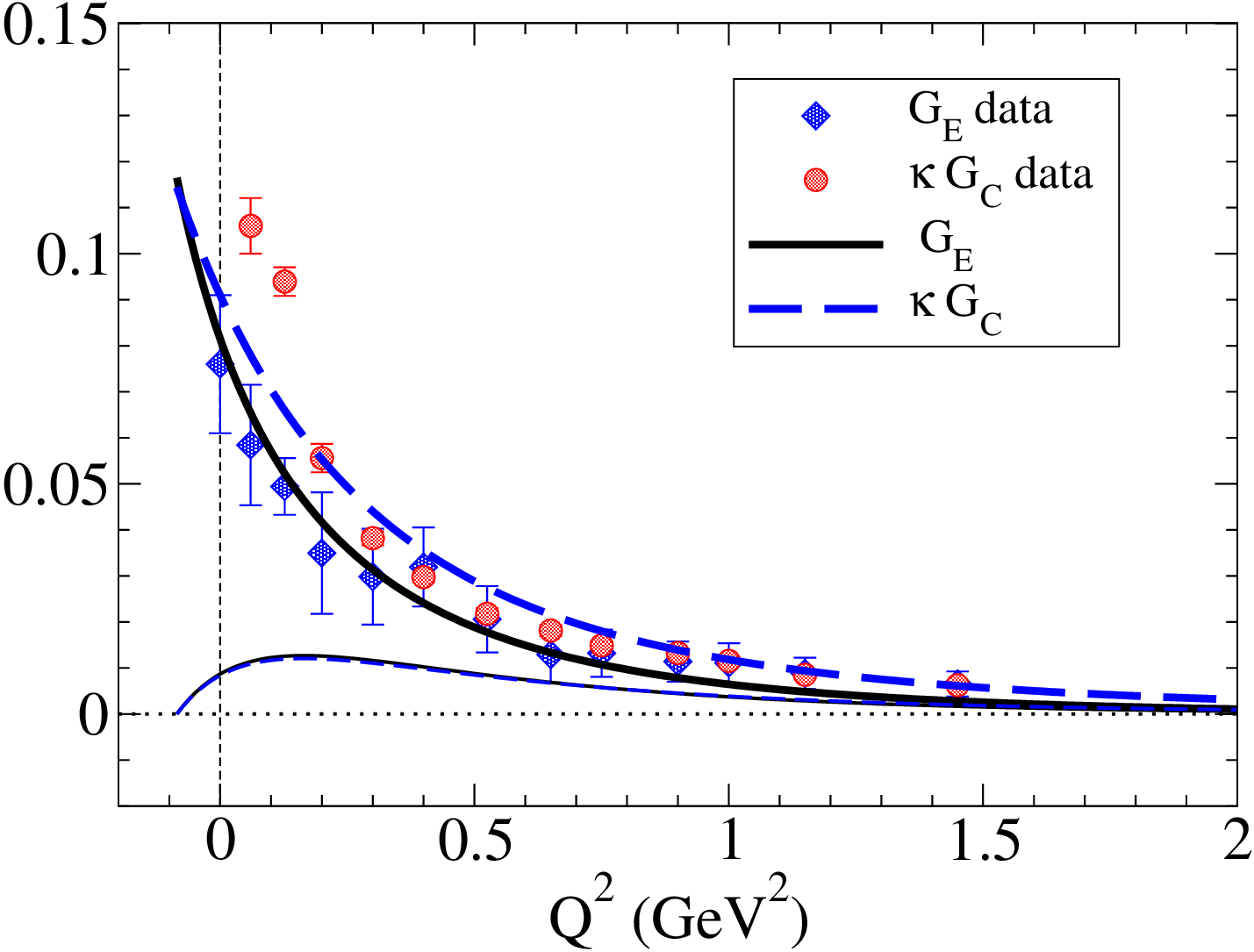}
\hspace{.6cm}
\includegraphics[width=7cm,clip]{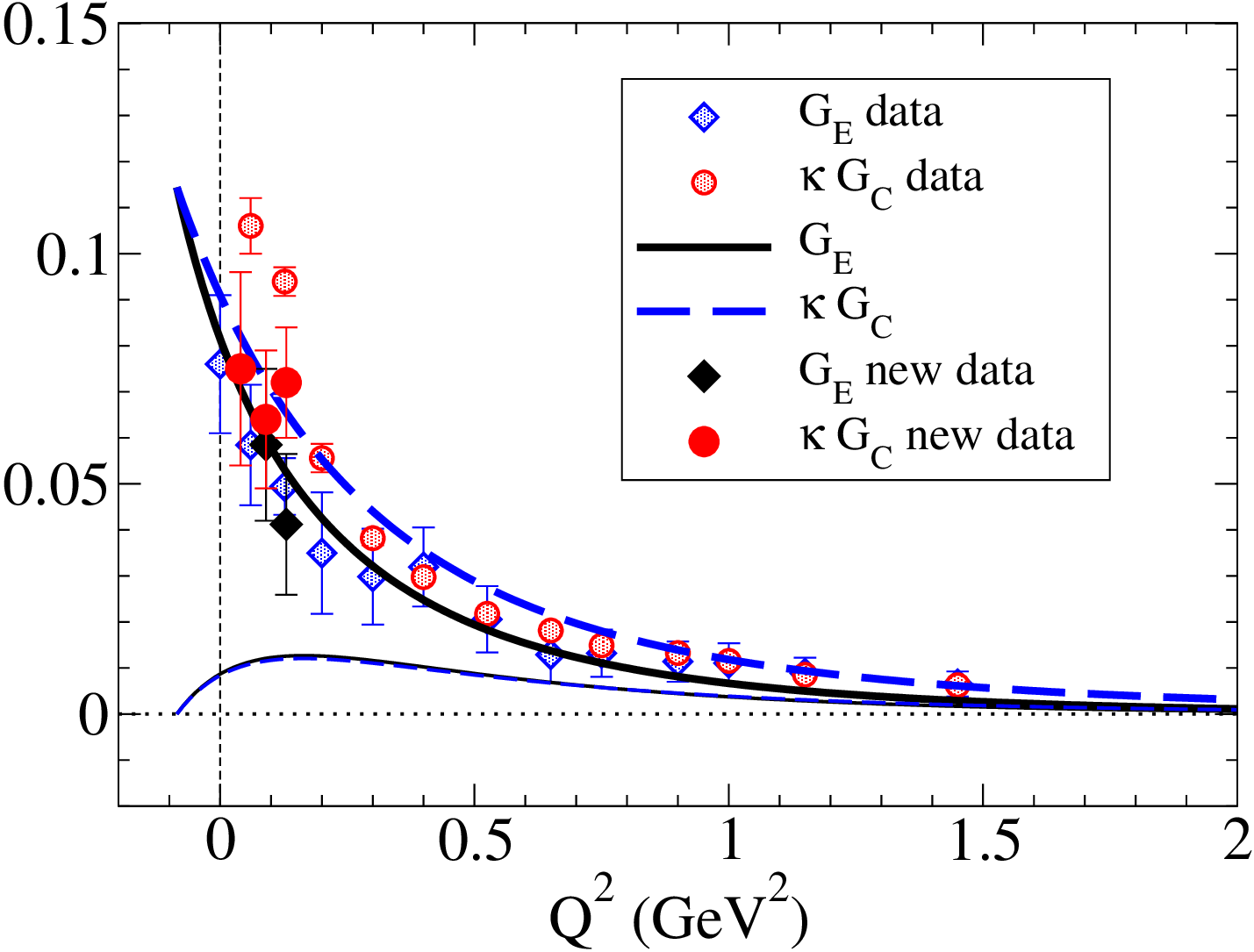}
\caption{$\gamma^\ast N \to \Delta(1232)$ quadrupole form factors.
The Coulomb form factor is normalized by 
$\kappa = \frac{M_\Delta-M_N}{2 M_\Delta}$ for convenience.
The model combines valence quark and pion cloud contributions.
Data from~\cite{MokeevDatabase,GlobalFit}.
{\bf Left:} 
Estimate with pion cloud contribution from~\cite{Siegert3}.
{\bf Right:}
Estimate with pion cloud contribution from~\cite{Siegert4}.
New low-$Q^2$ data from~\cite{Blomberg16a}.
}
\label{figDelta2}       % Give a unique label
\end{figure*}

%The covariant spectator quark model can 
%also be used to estimate transition form factors at low $Q^2$.
 
One can also use the covariant spectator quark model
to estimate transition form factors at low $Q^2$.
In general the estimates cannot be compared directly 
with the data because the meson cloud contributions 
can be significant and camouflage the valence quark contributions.

In the case of the $\gamma^\ast N \to \Delta(1232)$ 
there is the possibility of estimating the valence quark effects 
since there are lattice QCD data available~\cite{Alexandrou08}.
We obtain non-zero contributions to the 
electric ($G_E$) and Coulomb ($G_C$) 
quadrupole form factors
when we consider $d$-wave contributions to the 
$\Delta(1232)$ wave function~\cite{NDeltaD,LatticeD,DeltaDFF,DeltaShape}. 
The contributions of those states for the 
transition form factors are small
(about one order of magnitude) 
when compared with the experimental data, 
but their magnitude is comparable  with 
the lattice QCD data~\cite{NDeltaD,LatticeD}.
One can then use lattice QCD data to estimate
the free parameters of the quark model 
performing fits to the data~\cite{Lattice,LatticeD}.
Once extrapolated to the physical limit, one 
obtains the valence quark contributions to
the quadrupole form factors~\cite{LatticeD}.

The experimental data for $G_E$ and $G_C$ 
are represented in Fig.~\ref{figDelta2}.
For convenience of the discussion the results for $G_C$ 
are multiplied by $\kappa=  \frac{M_\Delta -M_N}{2 M_\Delta}$.
The valence quark contributions discussed above are 
represented by the thin lines 
(solid line for $G_E$ and dashed line for $\kappa G_C$).
Notice, in particular, the difference of magnitude 
in comparison with the final results (thick lines).
The thick lines are discussed next. 

% FIGURE 9

To describe the physical data, we need to take 
into account the pion cloud contributions to 
the electric and Coulomb quadrupole form factors.
There are in the literature 
estimates of the pion cloud contributions 
to the quadrupole form factors based 
on the large-$N_c$ limit~\cite{Pascalutsa07a,Buchmann09a}.
Those estimates are, however, inconsistent 
with the long wavelength limit also 
known as Siegert's theorem,
which states that in the limit $Q^2 = -(M_\Delta -M_N)^2$,
the two quadrupole form factors 
are related by~\cite{Jones73,Devenish76,Buchmann-ST,Siegert1,Siegert2,Siegert3,Siegert4,GlobalFit,LowQ2expansion}
\ba
G_E = \frac{M_\Delta -M_N}{2 M_\Delta} G_C.
\label{eqST}
\ea
The point $Q^2 = -(M_\Delta -M_N)^2$ is also known as pseudothreshold.

\begin{figure}[b]  %[t]
\centering
\includegraphics[width=8cm,clip]{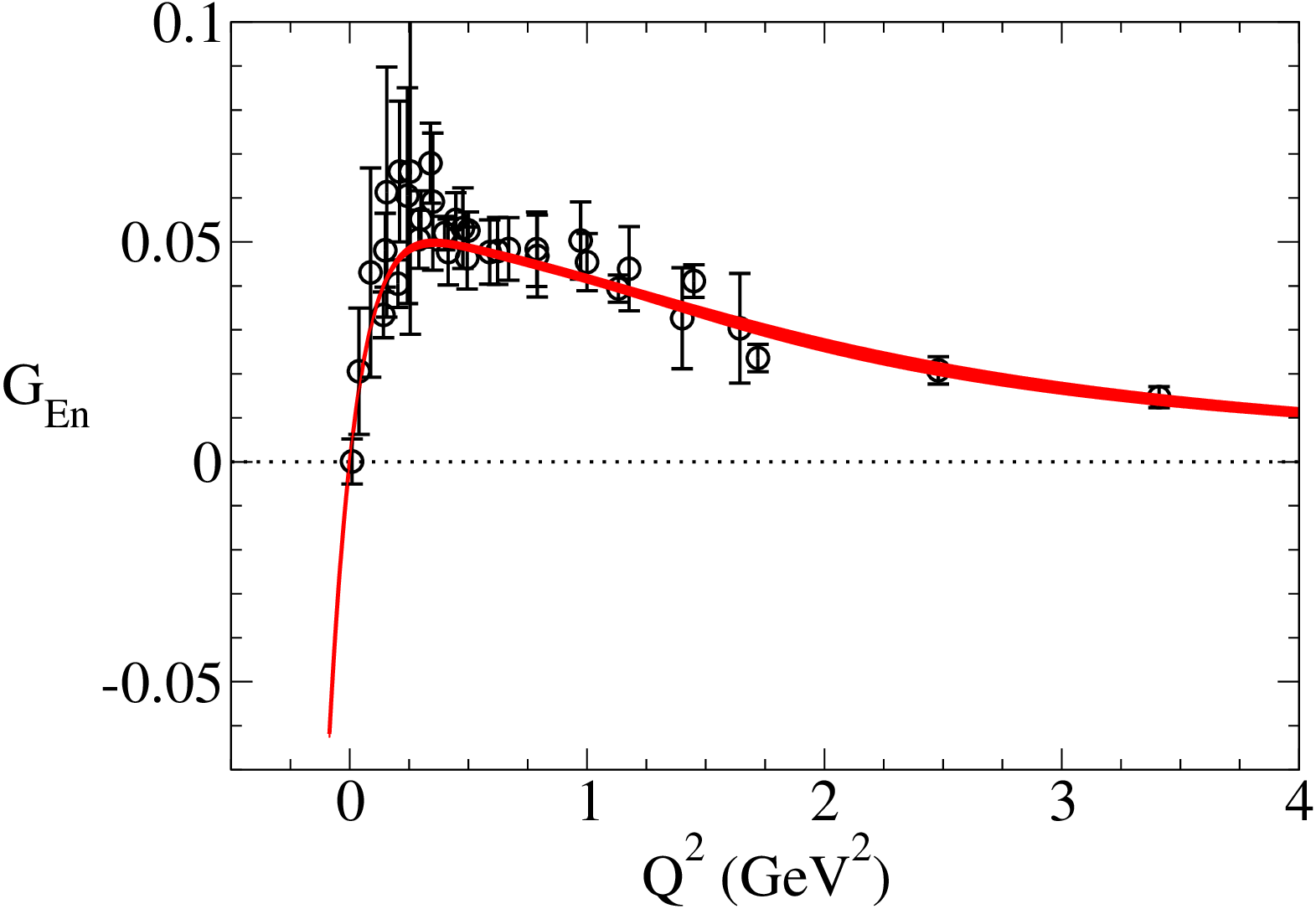}
\caption{Neutron electric form factor $G_{En}$
determined by a global fit to the 
$G_{En}$ and $G_E$ and $G_C$ quadrupole form factors~\cite{GlobalFit}.}
\label{figGlobalFit}       % Give a unique label
%\vspace{-.9cm}   %ZZZZ
\end{figure}

%\vspace{-2.4cm}

The original large-$N_c$ estimates of the pion cloud 
contributions~\cite{Pascalutsa07a,Buchmann09a,Siegert3}, 
can be improved with a correction 
of the order $1/N_c^2$ in $G_E$ in order to be consistent 
with Siegert's theorem (\ref{eqST})~\cite{Siegert3,Siegert4,GlobalFit}.
Combining the valence quark contributions 
with an improved large-$N_c$ estimate
of the pion cloud for $G_E$~\cite{Siegert3}, 
we obtain a good description of the quadrupole form factor data.
This can be observed in the left panel of the Fig.~\ref{figDelta2}.
The parametrization of~\cite{Siegert3} violates 
Siegert's theorem only a term in $1/N_c^4$.
The estimate fails only for the first two points for $G_C$.

On the right panel of Fig.~\ref{figDelta2},
we present a new calculation where the pion cloud 
parametrization is further improved in order to satisfy 
(\ref{eqST}) exactly.
In the graph, the low-$Q^2$ data for $G_E$ and $G_C$ are replaced by new data 
which corrects the previous analysis~\cite{Blomberg16a}.
The new estimate is in perfect agreement with the new data
for both form factors.

These results, demonstrate that 
the $\gamma^\ast N \to \Delta(1232)$ quadrupole form 
factors can be completely described by a combination 
of valence quark and pion cloud contributions.
The pion cloud parametrizations 
include no free parameters, and the valence quark contributions
are determined  exclusively by the lattice QCD data~\cite{LatticeD}.
There are, therefore, no fits to the physical data
in the estimates from Fig.~\ref{figDelta2}.

% FIGURE 9 

The pion cloud parametrizations for 
$G_E$ and $G_C$ are based on relations 
$G_E \propto G_{En}$ and $G_C \propto G_{En}$,
valid at low $Q^2$~\cite{Pascalutsa07a,Buchmann09a,Siegert3,Siegert4},
where $G_{En}$ is the neutron electric form factor.
Inspired by the previous results, 
we use the relation between the $\gamma^\ast N \to \Delta(1232)$ 
$G_E$ and $G_C$ form factors
and $G_{En}$ to obtain a more accurate 
parametrization of  $G_{En}$~\cite{GlobalFit}.
Our final parametrization for $G_{En}$
is presented in Fig.~\ref{figGlobalFit}.
Also estimated are the second moment of 
the neutron electric form factor:
$r_4^4 \simeq -0.4$ fm$^4$, 
and the electric and Coulomb square radii: 
$r_{E2}^2 = 2.2 \pm 0.2$ fm$^2$ and 
$r_{C2}^2 = 1.8 \pm 0.1$ fm$^2$.

% FIGURE 9

\section{Summary and Conclusions}
\label{secConclusions}

We present covariant estimates of the 
$\gamma^\ast N \to N^\ast$ transition form factors 
for large $Q^2$ for the $N^\ast$ states:
$\Delta(1232)$, $N(1440)$, $N(1520)$, $N(1535)$, 
$N(1880)$, $N(1700)$ and $\Delta(1700)$.
There are also estimates for the states 
$N(1650)$, $\Delta(1600)$ and $\Delta(1620)$,
not discussed in detail due to the lack of data for $Q^2> 2$ GeV$^2$.

Our estimates are valid in principle for $Q^2> 2$ GeV$^2$, 
since the valence quark effects are expected to dominate 
over the meson cloud effects when $Q^2$ increases. 
In some cases, however, we conclude 
that the valence quark contributions are insufficient 
to explain the data in the range $Q^2=1$--4 GeV$^2$,
implying that the meson cloud effects 
may be significant in that range.
Examples of those cases are the $\gamma^\ast N \to N(1535)$
Pauli form factor and the $\gamma^\ast N \to \Delta(1232)$
quadrupole form factors.

Our estimates are based mainly on the 
calibrations for the nucleon and $\Delta(1232)$ systems.
Future large $Q^2$ data, such as the data from the JLab-12 GeV upgrade 
will be fundamental to test our estimates 
at large $Q^2$, or to refine our calibrations, 
if the new data for the nucleon and the  $\Delta(1232)$
deviates from the present trend.

Future data at low $Q^2$, 
particularly in the range \mbox{$Q^2=0$--0.3 GeV$^2$,}
will be also very important to establish 
the shape of the transition form factors near $Q^2=0$.
The $N^\ast$ contributions to the nucleon Compton scattering 
are particularly sensitive to those form factors~\cite{Compton}.

%\clearpage

%\newpage

%\begin{acknowledgments}

\subsection*{Acknowledgments}
G.~R.~was supported by the Funda\c{c}\~ao de Amparo \`a
Pesquisa do Estado de S\~ao Paulo (FAPESP):
Project No.~2017/02684-5, Grant No.~2017/17020-BCO-JP.
%\end{acknowledgments}

%\clearpage

%\input{biblo}
%\vspace{-.5cm}

\end{document}